\documentclass[nofootinbib,prd,twocolumn,showpacs,showkeys,preprintnumbers]{revtex4-1}
\usepackage{hyperref,amssymb,amsmath,mathrsfs,bm,graphicx}
\usepackage{multirow}
\usepackage{array}
\usepackage{dblfloatfix}
\usepackage{dblfloatfix}
\usepackage{placeins}
\usepackage{float}
\usepackage[dvipsnames]{xcolor}
\begin{document}

\title {Integration of the Lane-Emden equation for relativistic anisotropic polytropes through Gravitational Decoupling: a novel approach.}
\author{D. Santana}
\affiliation{Departamento de F\'isica, Colegio de Ciencias e Ingenier\'ia, Universidad San Francisco de Quito,  Quito, Ecuador.\\}
\author{E. Fuenmayor }
\email{ernesto.fuenmayor@ciens.ucv.ve}
\affiliation{Centro de Física Teórica y Computacional, Escuela de Física, Facultad de Ciencias, Universidad Central de Venezuela, Caracas 1050, Venezuela.\\}
\author{E. Contreras }
\email{econtreras@usfq.edu.ec}
\affiliation{Departamento de F\'isica, Colegio de Ciencias e Ingenier\'ia, Universidad San Francisco de Quito,  Quito 170901, Ecuador.\\}

\begin{abstract}
In this work we propose a novel approach to integrate the Lane-Emden equations for relativistic anisotropic polytropes. We take advantage of the fact that Gravitational Decoupling allows to decrease the number of degrees of freedom once a well known solution of the Einstein field equations is provided as a seed  so after demanding the polytropic equation for the radial pressure the system is automatically closed. The approach not only allows to extend both isotropic or anisotropic known solutions but simplifies the computation of the Tolman mass whenever the Minimal Geometric Deformation is considered given that the $g_{tt}$ component of the metric remains unchanged. We illustrate how the the method works by analyzing
the solutions obtained from Tolman IV, Durgapal IV and Wymann IIa isotropic systems as a seed for the integration.
\end{abstract}
\maketitle
\section{Introduction}\label{intro}

Polytropic equations of state have played a long and remarkable role in astrophysics (see \cite{1pol,3pol} and references therein), and have been extensively used to study the stellar structure under a variety of fundamental astrophysical problems. For example, anisotropic white dwarfs have been modeled considering a general formalism to study Newtonian polytropes for anisotropic matter \cite{LHN, abellan20}, while for more compact configurations (e.g. neutron stars, super Chandrasekhar white dwarfs) see \cite{herrera2013, prnh, 6p}. In the general relativistic regime, polytropes also have been extensively studied (see, for example, \cite{4c,5, 6, 7, 8, 9,10, EF-K} and the references therein. For more recent works see \cite{Khan:2022ihg,Azam:2022qqj,Suarez-Urango:2021mjg}, for example).

The theory of polytropes is based on the polytropic equation of state, which can be written, in the case of a Newtonian-isotropic fluid like,
\begin{equation}
P=K\rho_0^{\gamma}=K\rho_{0}^{1+1/n} ,\label{Pola}
\end{equation}
where $P$ denotes the isotropic pressure,
$\rho_{0}$ stands for the mass (baryonic) density and $K$, $\gamma$, and $n$ are usually called  the polytropic constant, polytropic exponent, and polytropic index, respectively. Once the equation of state (\ref{Pola}) is assumed, the whole system is described by the Lane--Emden equation that may be numerically solved for any set of the parameters of the theory. When the constant $K$ is calculated from natural constants, the polytropic equation of state may be used to model a completely degenerate Fermi gas in the non-relativistic ($n = 5/3$) and relativistic limit ($n = 4/3$). In this case, Eq. (\ref{Pola}) provides
a way of modeling compact objects such as white dwarfs and allows   to obtain in a rather direct way the Chandrasekhar mass limit. Otherwise, if $K$ is a free parameter, the models can be used to describe either an isothermal ideal gas or a completely convective star. These models related to isothermal ideal gas are relevant in the so-called Sch\"onberg--Chandrasekhar limit \cite{3pol}.

Although local pressure isotropy is a very common assumption in the study of stellar objects, it is well known that many physical processes can produce deviations of the isotropy and/or fluctuations of the local anisotropy in pressure. These facts may be caused by a large variety of physical phenomena of the kind we expect to find especially in compact objects, so we find strong evidence that suggests that for certain ranges of density, a large number of physical phenomena can cause local anisotropy and therefore we must take it into account to describe realistic models. Among all the possibilities we would like to mention a few which might be particularly related to our primary interest. A possible source of anisotropy is related to intense magnetic fields observed in compact objects such as white dwarfs, neutron stars or magnetized strange quark stars \cite{12,13,15}. Another source constitutes the high indexes of viscosity expected to be present in neutron stars, in highly dense matter produced by opacity of matter to neutrinos in the collapse of compact objects \cite{16,18}, and the superposition of two isotropic fluids. It is important to note that, although the degree of anisotropy may be small, the effects produced on compact stellar objects may be appreciable \cite{herrera97,pr10,Ovalle:2016pwp,Ovalle:2018vmg}. So, the assumption of an isotropic pressure is a very stringent condition, specially in a situation in which the compact object is modeled as a structure with high density (as neutrons stars, for example). Besides, the isotropic pressure condition becomes unstable by the presence of physical factors such as dissipation, energy density inhomogeneity and shear as it has been recently proven in  \cite{LHP}. These facts explain the renewed interest in the study of fluids not satisfying the isotropic pressure condition and justify the interest of extending the theory of polytropes to anisotropic compact fluids. 

Thus, after assuming that the fluid pressure is anisotropic, the two principal stresses (say $P_r$ and $P_\bot$) are unequal and the polytrope equation of state reads
\begin{equation}P_r=K\rho_{0}^{\gamma}=K\rho_0^{1+1/n} .\label{Pol}
\end{equation}
As it happens in the Newtonian case, the introduction of the tangential pressure $P_{\perp}$ leads also in the relativistic case to an underdetermination of the problem requiring to impose an additional condition. Thus, in order to decrease the number of degrees of freedom the introduction of an additional condition is mandatory. In this regard, we can impose certain conditions on the metric variables as the vanishing of the Weyl tensor \cite{Herrera2001}, implemented in \cite{prnh} to obtain the conformally flat polytrope for anisotropic matter. This condition has its own interest, since it has been seen that highly compact configurations may be obtained with the specific distribution of anisotropy created by such a condition \cite{Herrera2001}. Other approaches as the Randall-Sundrum model \cite{RS} or $5$-dimensional warped geometries have served as an inspiration for other type of conditions, relating radial derivatives of the metric functions in spherically symmetric spacetimes, that produce self--gravitating spheres embedded in a $5$-dimensional flat space-time (embedding class one). Even more, models embedded in five dimensional spacetimes satisfy the so--called Karmarkar or class I condition \cite{karmarkar} (for recent developments see, \cite{EF-K,tello2,tello3,tello4,Nunez,tello5,maurya,Maurya:2020djz,kk2,kk4,kk5,kk9,kk10,kk12,kk13}, for example) which allows to choose one of the metric functions as the one which generates the total solution. We can also consider an extra equation of the state as polytropic equation of state for the tangential pressure as reported in \cite{Azam:2022qqj,Abellan:2020nkl}. Alternatively, the use of a new concept of complexity based on the scalar $Y_{TF}$ appearing in the orthogonal splitting of the Riemann tensor has increased in recent years \cite{VC}. It is important to mention that, very recent studies have considered stellar anisotropic models with a certain complexity (or fulfilling the vanishing complexity condition $Y_{TF}=0$) and it has been established a relationship between families of solutions that have different complexities with the possible occurrence of cracking \cite{GPRD, leon2021gravitational}.\\
In this work we take an alternative route to integrate the Lane-Emden equation which consists of considering stellar interiors supported by anisotropic fluids fulfilling the polytropic equation of state for the total obtained radial pressure in combination with the Gravitational Decoupling (GD) \cite{ovalle2017} through the Minimal Geometric Deformation approach (MGD) (see
\cite{ovalle2015,ovalle2018,estrada2018,ovalle2018a,lasheras2018,estrada,rincon2018,ovalleplb,tello2019,lh2019,estrada2019,gabbanelli2019,sudipta2019,leon2019,tello2019c,tello20,jorgeLibro,Ovalle:2020kpd,Contreras:2021yxe,contreras-kds,tello2021w,darocha1,darocha2,darocha3, sultana, maurya2021, neeraj, zubair, maurya2021a,Meert:2021khi,daRocha:2021sqd,Azmat:2022bbc,Ovalle:2022eqb,Heras:2022aho} for recent developments). It is worth mentioning that the GD has been used in order to extend know solutions of Einstein field equations by coupling different sources, to decoupling a complex energy momentum tensor in simpler components, to find new solutions in theories beyond general relativity, and to find both static and stationary black hole solutions.
In this work, we have demanded that the interaction between both the perfect (``seed'') and the decoupling fluids (the $\theta$ extra sector) is such that the total effective source fulfills a polytropic equation of state. 
As we shall explain later, the use of GD by MGD allows to extend well--known solutions (the so--called ``seed'' sector) by deforming the seed metric and adding an unknown source to the matter sector in a suitable manner. The advantage of such a procedure is that instead of providing two conditions to close the system of three differential equations in the static and spherically symmetric case, only one extra condition is required (because from the start it must be provided the seed solution). This procedures is clearly convenient in the integration of the relativistic Lane-Emden equation in the sense that it is sufficient with supplying the polytropic equation of state for the radial pressure to close the Einstein's system of field equations. More precisely, it is not necessary to propose any ansatz for the anisotropy or give some geometric condition but only a well--known stellar interior configuration.\\
This work is organized as follows. In the next section we study the basic equations of general relativity as well as a summary of the theory of relativistic polytropes. In Section \ref{GD}, we review the main aspects of GD through the Minimal Geometric Deformation (MGD) formalism. We dedicate Section \ref{Lane-Emden-GD} to obtaining the Lane-Emden equations for polytropes by gravitational decoupling and study some specific models. Finally, the last section is devoted to final remarks and conclusions.


\section{The General Relativistic polytrope for anisotropic matter}\label{polyaniso}

\subsection{The field equations and conventions}

Let us consider a static and spherically symmetric distribution of anisotropic matter which metric, in \break Schwarzschild--like coordinates, is parametrized as
\begin{eqnarray}\label{metric1}
ds^{2}=e^{\nu}dt^{2}-e^{\lambda}dr^{2}-r^{2}(
d\theta^{2}+\sin^{2}d\phi^{2}),
\end{eqnarray}
where $\nu$ and $\lambda$ are functions of $r$. The metric (\ref{metric1}) has to satisfy Einstein field equations \footnote{We are assuming natural units $G=c=1$}
\begin{eqnarray}\label{EE}
G^\mu_\nu = - 8 \pi T^\mu_\nu.
\end{eqnarray}

The matter content of the system (describing an anisotropic fluid) is represented by the energy--momentum tensor 
\begin{eqnarray}\label{tmunu1}
T_{\mu\nu}=(\rho+P_{\perp})u_{\mu}u_{\nu}-P_{\perp}g_{\mu\nu}+(P_{r}-P_{\perp})s_{\mu}s_{\nu},
\end{eqnarray}
where $\rho$ is the total energy density,
\begin{eqnarray}\label{v1}
u^{\mu}=(e^{-\nu/2},0,0,0),
\end{eqnarray}
is the four velocity
of the fluid which satisfies $u_{\mu}u^{\mu}=1$,
and $s^{\mu}$ is defined as
\begin{eqnarray}\label{v2}
s^{\mu}=(0,e^{-\lambda},0,0),
\end{eqnarray}
with the properties $s^{\mu}u_{\mu}=0$, $s^{\mu}s_{\mu}=-1$. 

Replacing (\ref{metric1}), (\ref{tmunu1}), (\ref{v1}) and (\ref{v2}) in (\ref{EE}), we obtain
\begin{eqnarray}
T^0_0=\rho&=&-\frac{1}{8\pi}\bigg[-\frac{1}{r^{2}}+e^{-\lambda}\left(\frac{1}{r^{2}}-\frac{\lambda'}{r}\right) \bigg],\label{ee1}\\
- T^1_1 = P_{r}&=&-\frac{1}{8\pi}\bigg[\frac{1}{r^{2}}-e^{-\lambda}\left(
\frac{1}{r^{2}}+\frac{\nu'}{r}\right)\bigg],\label{ee2}
\end{eqnarray}
\begin{equation}
-T^2_2= P_{\perp}=\frac{1}{8\pi}\bigg[ \frac{e^{-\lambda}}{4}
\left(2\nu'' +\nu'^{2}-\lambda'\nu'+2\frac{\nu'-\lambda'}{r}
\right)\bigg]\label{ee3},
\end{equation}
where primes denote derivative with respect to $r$. 

Furthermore, we shall consider that outside the fluid distribution the space--time is given by the Schwarzschild solution, namely
\begin{eqnarray}
ds^{2}&=&\left(1-\frac{2M}{r}\right)dt^{2}-\left(1-\frac{2M}{r}\right)^{-1}dr^{2}\nonumber\\
&&-r^{2}(d\theta^{2}+\sin^{2}d\phi^{2}),
\end{eqnarray}
where $M$ represents the total energy of the system.
In order to match the two metrics smoothly on the boundary surface $r=r_{\Sigma}=\rm constant$, we require continuity of the first and second fundamental forms across that surface. As a result of this matching we obtain the well known result,
\begin{eqnarray}
e^{\nu_{\Sigma}}&=&1-\frac{2M}{r_{\Sigma}},\label{nursig}\\
e^{-\lambda_{\Sigma}}&=&1-\frac{2M}{r_{\Sigma}},\label{lamrsig}\\
P_{r_{\Sigma}}&=&0,\label{prr}
\end{eqnarray}
where the subscript $\Sigma$ indicates that the quantity is evaluated at the boundary surface.

From the radial component of the conservation law,
\begin{eqnarray}\label{Dtmunu}
\nabla_{\mu}T^{\mu\nu}=0,
\end{eqnarray}
one obtains  the generalized Tolman--Oppenhei-\break mer--Volkoff equation for anisotropic matter  which reads,
\begin{eqnarray}\label{TOV}
P_{r}'=-\frac{\nu'}{2}(\rho +P_{r})+\frac{2}{r}(P_{\perp}-P_{r}).
\end{eqnarray}
Alternatively, using
\begin{eqnarray}
\nu'=2\frac{m+4\pi P_{r}r^{3}}{r(r-2m)},
\end{eqnarray}
where the mass function $m$ is as usually defined by
\begin{eqnarray}\label{m}
e^{-\lambda}=1-2m/r ,
\end{eqnarray}
we may rewrite Eq. (\ref{TOV}) in the form
\begin{eqnarray}\label{TOV2}
P_{r}'=-\frac{m+4\pi r^{3} P_{r}}{r(r-2m)}(\rho+P_{r})+\frac{2}{r}\Delta, \label{TOVb}
\end{eqnarray}
where
\begin{eqnarray}\label{anisotropy}
\Delta=P_{\perp}-P_{r},
\end{eqnarray}
measures  the anisotropy of the system.

For the physical variables appearing in (\ref{TOVb}) the following boundary conditions apply
\begin{eqnarray}
 m(0)=0,\qquad m(r_{\Sigma})=M, \qquad P_{r}(r_{\Sigma})=0,
\end{eqnarray}
which corresponds to a stellar configuration surrounded by the Schwarzschild vacuum \cite{sch}
As already mentioned in the introduction, in order to integrate equation (\ref{TOVb}), we shall need additional information. In this work we propose a novel scenario; namely, we satisfy Einstein's system of field equations using the polytropic equation of state for the radial pressure together with the Gravitational Decoupling \cite{ovalle2017}, by means of the Minimal Geometric Deformation approach (once the seed solution is given). In the next section we will implement the polytropic equation of state to construct the generalized anisotropic Lane--Emden equation which will arrive after introducing the polytropic equation of state in the TOV equation given by (\ref{TOV2}) followed by some redefinitions of the parameters involved.

\subsection{Relativistic anisotropic polytropes}

When considering the polytropic equation of state within the context of general relativity two different possibilities arise leading to the same equation in the Newtonian limit \cite{herrera2013}. The first one, preserves the original polytropic equation of state (\ref{Pola}) and the second case allows another (natural) possibility that consists in assuming that the relativistic polytrope is defined by,
\begin{eqnarray}
P_{r}=K\rho^{\gamma}=
K\rho^{1+\frac{1}{n}} \label{pr1b}.
\end{eqnarray}
In this case the baryonic density $\rho_0$ is replaced by the total energy density $\rho$ in the polytropic equation of state. The general treatment is very similar for both cases and therefore, for simplicity, we shall restrict here to the case described by (\ref{pr1b}). It can be shown that the relationship between the two densities is given by \cite{herrera2013},
\begin{eqnarray}
\rho= \frac{\rho_0}{\left(1 - K \rho_0^{1/n}\right)^n}
\label{Rho}.
\end{eqnarray}

As it is well known from the general theory of polytropes, there is a bifurcation at the value $\gamma=1$. Thus, the cases $\gamma=1$ and $\gamma\neq1$ have to be considered separately. In the context of our work we want to focus in bounded (fluid) compact star models so the $\gamma=1$ will not be considered. 

Let us define the variable $\psi$ by
\begin{eqnarray}\label{rho1}
\rho = \rho_c \psi^n,
\end{eqnarray}
where $\rho_{c}$  denotes the energy density at the center (from now on the subscript $c$ indicates that the variable is evaluated at the center). Now, we may rewrite (\ref{pr1b}) as
\begin{eqnarray}\label{Pr}
 P_{r} = K \rho^{\gamma} = K \rho_c^{\gamma} \psi^{n+1} = P_{rc} \psi^{n+1},
\end{eqnarray}
with $P_{rc}=K\rho_{c}^{\gamma}$. Replacing (\ref{pr1b}) and (\ref{rho1}) in (\ref{TOV2}), the TOV equation can be written as
\begin{eqnarray}\label{lmeq}
(n+1)  \psi' &=& - \left(\frac{m + 4 \pi P_{rc} \psi^{n+1}  r^3}{r(r -2m)} \right) \frac{1}{\alpha} \left( 1 + \alpha \ \psi^{n} \right)\nonumber\\
&&\hspace{4.2cm}+ \frac{2 \Delta}{r P_{rc}},
\end{eqnarray}
where $\alpha=P_{rc}/\rho_{c}$.

Let us now introduce the following dimensionless variables
\begin{eqnarray}
     r &=&\frac{\xi}{A}\label{eq:var0},\\ 
     A^2 &=& \frac{4 \pi \rho_c}{\alpha(n+1)} \label{eq:var1}, \\
    \psi^n &=& \frac{\rho}{\rho_c},\\
    \eta(\xi) &=& \frac{m(\xi) A^3}{4 \pi \rho_c},  \label{eq:var2}
\end{eqnarray}
from where (\ref{lmeq}) reads (see \cite{herrera2013, 6p} for details)
\begin{eqnarray}\label{lemd}
&&\xi^2 \frac{d\psi}{d\xi} \left[ \frac{1 - 2 \alpha (n+1) \frac{\eta}{\xi}}{1 + \alpha \ \psi } \right]+ \eta + \alpha \xi^3 \psi^{n+1}\nonumber\\
&& - \frac{2 \Delta \ \xi}{P_{rc} \psi^n (n+1)} \left[ \frac{1 - 2 \alpha (n+1) \frac{\eta}{\xi}}{1 + \alpha \ \psi } \right] = 0,
\end{eqnarray}
with 
\begin{eqnarray}\label{masseta}
\eta'= \xi^{2}\psi^{n}.
\end{eqnarray}
Equation (\ref{lemd}) in combination with 
(\ref{masseta})
corresponds to the generalized Lane--Emden equation for an anisotropic fluid characterized by a polytropic equation of state.


\section{Gravitational decoupling}\label{GD}

In this section we introduce the GD by MGD (for more details, see \cite{ovalle2017}). Let us start by considering the Einstein field equations (\ref{EE}) sourced by certain $T_{\mu\nu}^{(tot)}$ which can be written as
\begin{equation}\label{energy-momentum}
   T_{\mu\nu}^{(tot)} = T^{(s)}_{\mu\nu} + \beta\theta_{\mu\nu}\;,
\end{equation}
where $T^{(s)}_{\mu\nu}$ represents the matter content of a known solution, namely the {\it seed} sector,  and $\theta_{\mu\nu}$ describes an extra source coupled through the parameter $\beta$. It is essential to point out that the additional term $\beta \Theta_{\mu \nu}$ is not considered a perturbation, i.e., the coupling parameter $\beta$ could indeed be larger than unity (such coupling is introduced in order to control the effect of the unknown anisotropic source). Note that, since the Einstein tensor fulfills the Bianchi's identities, the total energy--momentum tensor  satisfies the conservation equation
\begin{equation}\label{divergencia-cero-total}
    \nabla_{\mu} T^{\mu\nu} = 0\;.
\end{equation}
so that, whenever 
$\nabla_\mu T^{\mu\nu(s)} = 0$, the condition
\begin{equation}\label{divergencia-cero-theta}
    \nabla_\mu \theta^{\mu\nu} = 0\;,
\end{equation}
is automatic and as a consequence, there is no exchange of energy-momentum between the seed solution and the extra source $\theta^{\mu\nu}$ (the interaction is entirely gravitational).\\

We are restricting ourselves to the spherically symmetric static anisotropic fluid case with internal metric (in Schwarzschild coordinates) given in (\ref{metric1}). In this case the source can be expressed like,
\begin{eqnarray}
T^{\mu(s)}_{\nu}&=&\textnormal{diag}(\rho^{(s)},-P_{r}^{(s)},-P^{(s)}_{t},-P^{(s)}_{\perp}),\label{tmunu}\\
\theta^{\mu}_{\nu}&=&\textnormal{diag}(\theta_{0}^{0},\theta_{1}^{1},\theta_{2}^{2},\theta_{2}^{2}).\label{thetamunu}
\end{eqnarray}
Note that, given the symmetry of the system, the components of the extra source, $\theta_{\mu\nu}$, depend on the radial coordinate only. To be more precise, although we are assuming anisotropic systems, it means that both the radial and the tangential pressures are different so that $\theta_{11}\ne\theta_{22}=\theta_{33}$. Even more, although an explicit dependence on the angular coordinates could represent an anisotropic systems too, this break our assumption that the model we are considering here is a spherically symmetric one. 

Then, equations (\ref{EE}), (\ref{energy-momentum}), (\ref{tmunu}) and (\ref{thetamunu}) lead to the fact that the total energy-momentum tensor ($T_{\mu\nu}^{(tot)}$) satisfies the system of field equations 
(\ref{ee1}), (\ref{ee2}) and (\ref{ee3}), where now, in the left hand side we have
\begin{eqnarray}
    \rho &=& \rho^{(s)} + \beta\theta_{0}^{0} \;,\label{mgd07a}\\
P_{r} &=& P_r^{(s)}-\beta\theta_{1}^{1}   \;,\label{mgd07b}\\
    P_{\perp} &=& P_{\perp}^{(s)} -\beta\theta_{2}^{2} \;.\label{mgd07c}
\end{eqnarray}
Because, in general, $\theta^1_1 \neq \theta^2_2$, we find that the system represents an anisotropic fluid. Is clear that the non-linearity of Einstein’s equations avoids that the decomposition (\ref{energy-momentum}) leads to two sets of equations; one for each source involved. Nevertheless, the decoupling is possible in the context of MGD as we shall demonstrate in what follows.\\

Let us introduce a \textit{geometric deformation} in the metric functions given by 
\begin{eqnarray}
    e^{-\lambda}\;\; &\longrightarrow &\;\;e^{-\lambda}= e^{-\mu} + \beta f\;,\label{comp-radial}
\end{eqnarray}
where $f$ is the so-called decoupling function and $\beta$ is the same free parameter that ``controls'' the influence of $\theta_{\mu\nu}$ on $T_{\mu\nu}^{(s)}$. It is worth mentioning that although a general treatment considering deformation in both components of the metric is possible, in this work we shall concentrate in the particular case where the deformation is only implemented on the $g^{rr}$ component. Now, replacing (\ref{comp-radial}) in the system (\ref{ee1})-(\ref{ee3}), we are able to split the complete set of differential equations into two subsets: one describing a seed sector sourced by the conserved energy-momentum tensor $T_{\mu\nu}^{(s)}$,
    \begin{eqnarray}
     \rho^{(s)} &=& \frac{1}{8\pi}\left[\frac{1}{r^2} +
        e^{-\mu}\!\left(\frac{\mu'}{r} - \frac{1}{r^2}\right)\right]\!,\label{mgd13}
        \\
     P_r^{(s)} &=& \frac{1}{8\pi}\left[-\frac{1}{r^2} +
        e^{-\mu}\!\left(\frac{\nu'}{r} + \frac{1}{r^2}\right)\right]\!,\label{mgd14}
        \\
       P_{\perp}^{(s)} &=& \frac{1}{32\pi}e^{-\mu} \!\! \left( 2\nu'' + {\nu'}^2 
    - \mu' \nu' + 2\frac{\nu'-\mu'}{r} \right) \!,\nonumber\\\label{mgd15}
\end{eqnarray}
and the other set corresponding to quasi-Einstein field equations sourced by $\theta_{\mu\nu}$,
 \begin{eqnarray}
    \rho^{\theta} &=&- \frac{\beta}{8\pi}\left(\frac{f}{r^2}+ \frac{f'}{r}\right)\,,\label{mgd16}\\
    P_{r}^{\theta} &=& \frac{\beta f}{8\pi}
        \left(\frac{\nu'}{r} + \frac{1}{r^2}\right)\!,\label{mgd17}\\
    P_{\perp}^{\theta} &=& \frac{\beta}{8\pi}\left[\frac{f}{4} \left( 2\nu'' + 
    {\nu'}^2 + 2\frac{\nu'}{r}\right)
    +\frac{f'}{4} \left( \nu' + \frac{2}{r} \right)\right]\!,\nonumber\\\label{mgd18}
    \end{eqnarray}
where we have defined $\rho^{\theta}=\beta\theta^{0}_{0}$, 
$P_{r}^{\theta}=-\beta \theta^{1}_{1}$
and 
$P_{\perp}^{\theta}=-\beta\theta^{2}_{2}$.
Note that, as the seed sector is sourced by a conserved energy--momentum tensor, namely $\nabla_{\mu}T^{\mu\nu(s)}=0$, the the components of $\theta_{\mu\nu}$ satisfy the conservation equation $\nabla_{\mu}\theta^{\mu}_{\nu}=0$, given by
\begin{eqnarray}\label{consthe}
(P_{r}^{\theta})'+ \frac{\nu'}{2}(\rho^{\theta}+P_{r}^{\theta}) 
- \frac{2}{r}(P_{\perp}^{\theta}-P_{r}^{\theta})=0,
\end{eqnarray}
Remarkably, although the quasi–Einstein equations differ from Einstein equations, the expression given in (\ref{consthe}) is completely analogous to the anisotropic Tolman-Opphenheimer-Volkoff
(TOV) equation.
Next, to complete the process, we require to satisfy the matching conditions (\ref{nursig}), (\ref{lamrsig}) and (\ref{prr}) on the boundary surface $\Sigma$. \\

To conclude this section, we would like to emphasize the importance of GD  by MGD as a useful tool to find solutions of EFE. As it is well known, in static and spherically symmetric spacetimes sourced by anisotropic fluids, EFE reduce to three equations and five unknowns, namely $\{\nu,\lambda,\rho,P_{r},P_{\perp}\}$. In this sense, two auxiliary conditions must be provided: metric conditions, equations of state, complexity of the system, etc. However, in the context of MGD a seed solution is given, namely, a metric $\{\nu,\mu\}$ that solve Eqs. (\ref{mgd13})-(\ref{mgd15}). Now, note that as we are only deforming the radial metric by Eq. (\ref{comp-radial}) , the decoupling sector system given by Eqs. (\ref{mgd16})-(\ref{mgd18}) has a metric given by the pair $\{\nu,f\}$. In this regard, it worth emphasising that through the MGD the number of degrees of freedom automatically reduces from five to four, namely $\{f,\theta^{0}_{0},\theta^{1}_{1},\theta^{2}_{2}\}$, given that both the seed and the decoupling sector share the same temporal metric so that only one extra condition is required. In general, this condition is implemented in the decoupling sector given by Eqs. (\ref{mgd16}), (\ref{mgd17}) and (\ref{mgd18}) by some equation of state (or condition) which leads to a differential equation for the decoupling function $f$. For example:
i) The mimic constrain for the density, namely $\theta^{0}_{0}=\rho^{(s)}$, which ensures that both secctors (seed and decoupling) have the same energy density profile. ii) The mimic constrain for the pressure, namely $-\theta^{1}_{1}=P_{r}^{(s)}$, which ensures that both sector have the same profile for the radial pressure.
Another possibility is to supply a condition for the total solution (which of course entails a relation between the two sectors). For example, if we consider the simple barotropic condition $P_{r}=\rho$, this encodes the relation 
\begin{eqnarray}
P^{(s)}_{r}-\beta \theta^{1}_{1}=\rho^{(s)}+\beta \theta^{0}_{0}
\end{eqnarray}
so in the context of GD the system is closed enough. In this work we propose the polytropic equation of state
\begin{eqnarray}
P_{r}=K \rho^{\gamma},
\end{eqnarray}
which encodes the relation
\begin{eqnarray}
P^{(s)}_{r}-\beta \theta^{1}_{1}=K (\rho^{(s)}+\beta \theta^{0}_{0})^{\gamma}
\end{eqnarray}
between both sectors. As we shall see later, we demonstrate that this strategy allows to write the Lane--Emden equation and the
anisotropy present in the total solution in terms of the pair $(\eta,\xi)$, defined in (\ref{eq:var0}) and (\ref{eq:var2}), and the metric variables of the seed sector.


\section{Gravitational decoupling and anisotropic polytropes}\label{Lane-Emden-GD}

In this section we will outline the procedure to obtain the structural Lane-Emden equations by means of the GD. In order to do so, we note that integration of Eqs. (\ref{lemd}) and (\ref{masseta}) provides $(\psi,\eta)$ as a function of $\xi$ whenever the anisotropy
$\Delta=P_{\perp}-P_{r}$ is supplied. In this work we take advantages of the GD to write $\Delta$ in terms of $(\eta,\xi)$ and the seed metric functions $(\nu,\mu)$ as we shall explain in what follows. 

First, note that the total anisotropy can be written as
\begin{eqnarray}\label{anis}
\Delta=\Delta^{(s)}+\Delta^{\theta}
\end{eqnarray}
with $\Delta^{(s)}=P_{\perp}^{(s)}-P_{r}^{(s)}$ and
$\Delta^{\theta}=P_{\perp}^{\theta}-P_{r}^{\theta}$. Now, by 
replacing (\ref{mgd14}), (\ref{mgd15}), (\ref{mgd17}) and (\ref{mgd18}) in (\ref{anis}) we obtain
\begin{eqnarray}\label{delta}
    &&\Delta = \frac{A^2}{32\pi} \left[ e^{-\mu} (2\nu^{\prime \prime} + \nu^\prime - \mu^\prime \nu^\prime -2 \frac{\nu^\prime +\mu^\prime}{\xi} -\frac{4}{\xi^2}) + \frac{4}{\xi^2} \right]\nonumber\\
    &&\; + \frac{\beta}{32\pi} \left[ f A^2 \left( 2\nu^{\prime \prime} +\nu^{\prime^2} -2\frac{\nu^\prime}{\xi} - \frac{4}{\xi^2} \right) +f^\prime A \left( \nu^\prime + \frac{2}{\xi} \right) \right]\nonumber,\\
\end{eqnarray}
where, from now on, primes indicate derivation respect to the variable $\xi$. Next, from (\ref{m}) and (\ref{comp-radial}) and using (\ref{eq:var0}), (\ref{eq:var1}) and (\ref{eq:var2}), the decoupling function $f$ and its derivative can be written as
\begin{eqnarray}
f&=&\frac{1}{\beta}\left[ 1 - 2 \alpha (n+1) \frac{ \eta}{\xi} -e^{-\mu}  \right]\label{fxi},\\
f'&=&\frac{A}{\beta} \left[\mu^\prime e^{-\mu} +2\alpha (n+1)\frac{(\eta -\eta^\prime \xi )}{\xi^2}  \right],\label{fpxi}
\end{eqnarray}      
with $\eta$ given by (\ref{eq:var2}). Finally, replacing (\ref{fxi}) and (\ref{fpxi}) in (\ref{delta}) we arrive at
\begin{eqnarray}\label{anisotropia final}
    &&\Delta = \frac{\pi \rho_c}{\alpha (n+1)8\pi} \times \nonumber\\
    && \qquad\quad\quad \bigg[ \left(\mu^\prime e^{-\mu} +2\alpha (n+1)\frac{(\eta -\eta^\prime \xi )}{\xi^2}  \right)\Big( \nu^\prime + \frac{2}{\xi} \Big)\nonumber\\
                   &&+     \left( 1-2\alpha (n+1)\frac{\eta}{\xi} -e^{-\mu} \right) \left( 2\nu^{\prime \prime} +\nu^{\prime^2} -2\frac{\nu^\prime}{\xi} - \frac{4}{\xi^2} \right)\bigg],\nonumber\\
     \end{eqnarray}
where we have used the dimensionless variables defined by equations (\ref{eq:var0})-(\ref{eq:var2}). Note that, the final expression of the total anisotropy only depends on $(\eta,\xi)$ (dimensionless mass--function and radius) and the known metric functions of the seed sector $(\nu,\mu)$, as expected. At this point, it is worth emphasizing the system is closed in the following sense. As we stated before, for any anisotropic and spherically symmetric system, the problem reduces to solving three Einstein field equations for five unknowns, namely the metric functions $\{\nu,\lambda\}$ and the variable of the matter sector $\{\rho,P_{r},P_{\perp}\}$. In this regard, two additional conditions are required in order to close the system. In this work, one of this conditions is supplied by the polytropic equation of state relating the radial pressure with the energy density given by Eq. \ref{pr1b}. The other condition corresponds to that the temporal metric $\nu$ is a known function once a seed solution is specified.

To complement the discussion, we proceed to calculate the Tolman-Whittaker mass \cite{LHTolman}, $m_{T}$,  which in the context of GD is a simple task and this fact in itself reinforces the use of this tool novelty linked to the study of polytropes. Indeed, using the definition of the active gravitational mass,
\begin{equation}\label{TM}
    m_{T}=\frac{1}{2}e^{(\nu-\lambda)/2} \nu^\prime  r^{2},
\end{equation}
it is noted that the only relevant information that should be provided is the mass function that is encoded in $e^{-\lambda}$. This is so, given that the $g_{tt}$ metric component is known since it corresponds to the metric function of the seed solution. Thus, replacing (\ref{m}) in (\ref{TM}) and using (\ref{eq:var2}), we arrive at
 \begin{equation}
      m_T= \frac{1}{2} \left[e^\nu (1-\frac{2\eta \alpha (n+1)}{\xi})  \right]^{1/2} \nu^\prime \frac{\xi^2}{A^2},
  \end{equation}
  from where
  \begin{equation}
      \eta_T= \frac{\xi^2 (n+1) \nu^\prime}{2(4\pi \rho_c \alpha)^2} \left[   e^\nu \left(1-\frac{2\eta \alpha (n+1)}{\xi}\right) \right]^{1/2}
  \end{equation}
with
\begin{eqnarray}
\eta_T=\frac{m_T}{4\pi \rho_c \alpha^3}.
\end{eqnarray}
For the numerical calculations it is convenient to change to the following dimensionless variables
\begin{eqnarray}
    x &=& \frac{r}{r_\Sigma} = \frac{\xi}{r_\Sigma A}=\frac{\xi}{\xi_\Sigma},\\ 
        y &=& \frac{M}{r_\Sigma} = \alpha (n+1) \frac{\eta_\Sigma}{\xi_\Sigma}, \label{y}     
 \end{eqnarray}
 in terms of which the Tolman mass can be written as
   \begin{equation}
      \eta_T= \frac{\xi^2 (n+1) \nu^\prime}{2(4\pi \rho_c \alpha)^2} \left[   e^\nu \left(1-\frac{2 y \eta }{x \eta_\Sigma}\right) \right]^{1/2}.
  \end{equation}
We observe from (\ref{y}) that each anisotropic polytropic model is characterized by a unique $y$.

In what follows we shall proceed to integrate the Lane-Emden equations by considering different seed sectors and be able to obtain, through the MGD formalism, relativistic anisotropic polytropes as the total solution. To achieve this, we will introduce the final anisotropy function given in (\ref{anisotropia final}) into the generalized TOV or Lane-Emden equation for an anisotropic fluid characterized by a polytropic equation, given by (\ref{lemd}).


\subsection{Tolman IV}

In this section we shall use the well known Tolman IV solution \cite{tolman1939} as a seed so we can close the system and find an anisotropic total solution that also complies with the fact of satisfying a polytrope equation of state, this, carried out within the framework of GD by means of MGD, as was previously explained. Then, we have
\begin{eqnarray}
e^{\nu}&=&b^2 \left( 1+\frac{r^2}{a^2} \right)\\
e^{-\mu}&=&\frac{(c^2-r^2)(a^2+r^2)}{c^2(a^2+2r^2)},
\end{eqnarray}
where $a$, $b$ and $c$ are constants, that have the same dimensions of $r$. In order to describe our model, we have fixed the values of $a=1.41421$, $b=2.23607$ (in dimensions of a length) after a numerical probing in the parameters space.

\begin{figure}[ht!]
		\centering
		\includegraphics[width=0.5\textwidth]{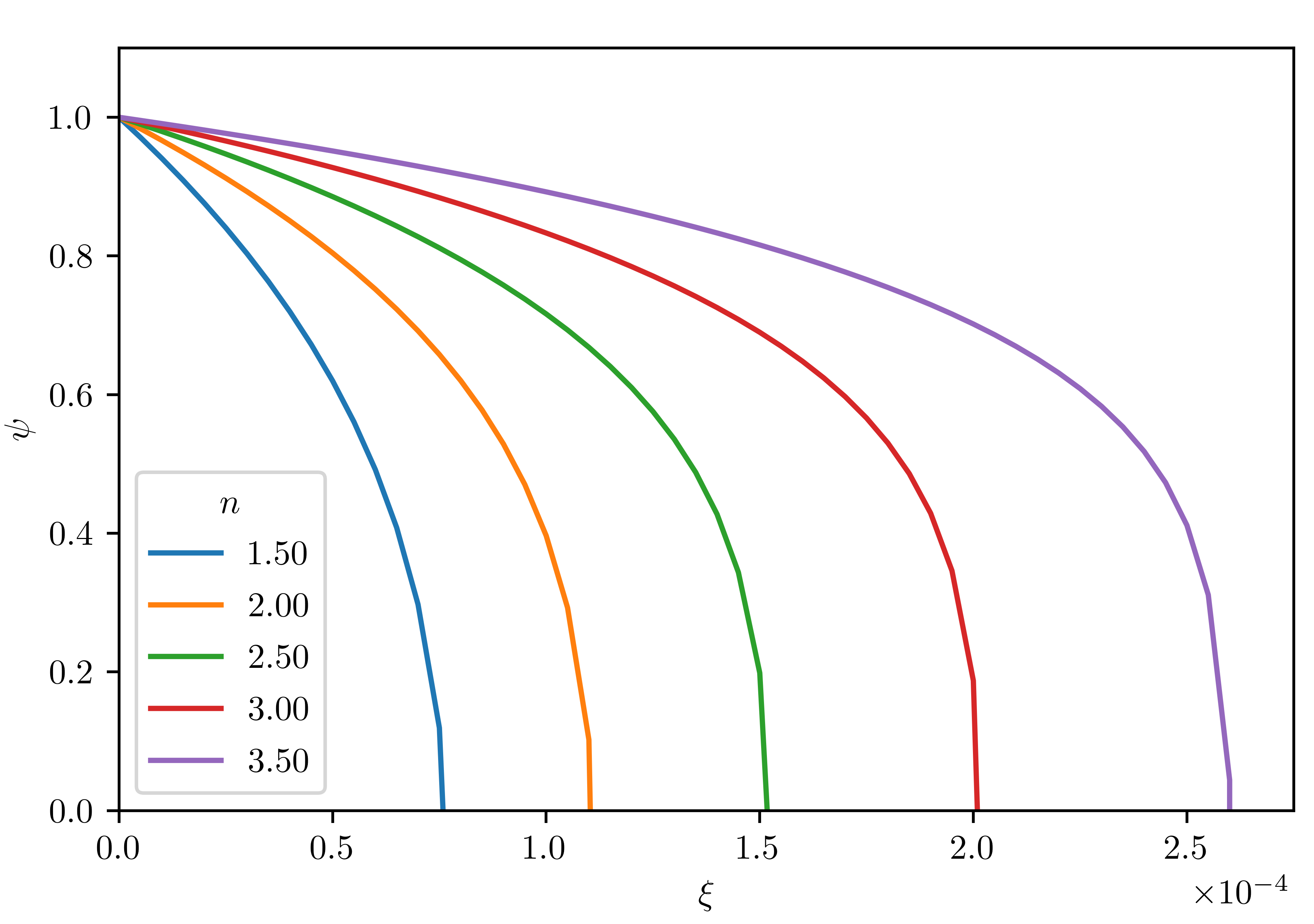}
		\caption{$\psi$ as a function of $\xi$ for $\alpha=0.5$ and different values of $n$.}
		\label{fig: tolman-psi}
\end{figure}

In Fig. \ref{fig: tolman-psi} we perform the integration of system of equations (\ref{lemd}) and (\ref{masseta}) for the
values of the parameters indicated in the figure legend where we represent $\psi$ (energy density) as a function of the dimensionless variable $\xi$ for different values of the polytropic index $n$ for the polytrope model obtained with the Tolman IV seed-solution. We observe that $\psi$ is monotonously decreasing, as expected for well behaved general relativistic polytropes and the radial pressure $P_r$ vanishes at the surface as required
by the continuity of the second fundamental form. 

\begin{figure}[ht!]
		\centering
		\includegraphics[width=0.5\textwidth]{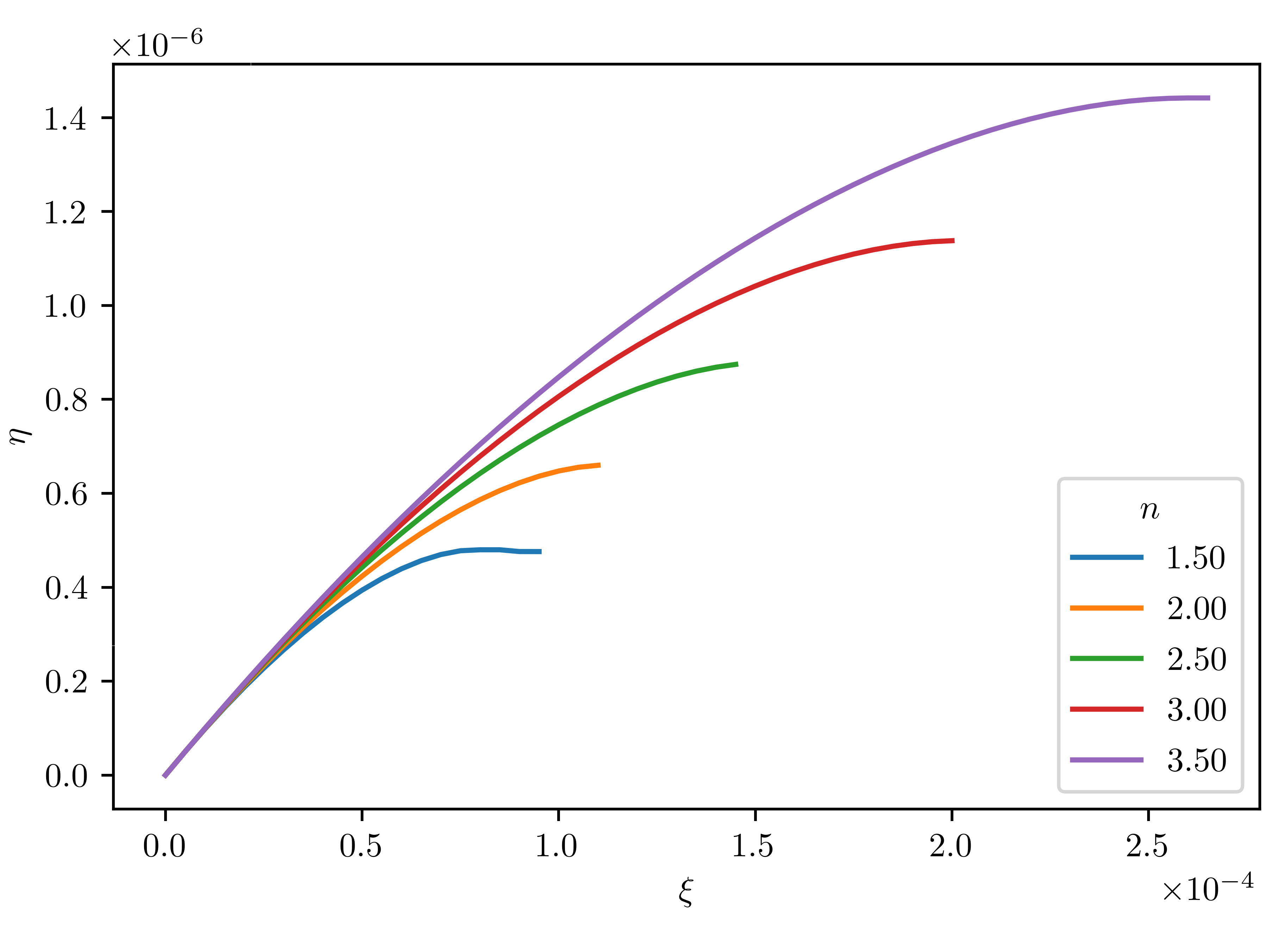}
		\caption{$\eta$ as a function of $\xi$ for $\alpha=0.5$ and different values of $n$.}
		\label{fig:tolman-eta}
\end{figure}

In Fig. \ref{fig:tolman-eta} it is shown the mass-function $\eta$ as a function of the variable $\xi$ for different values of the index $n$. The results also depends on the ``rigidity parameter'' ($\alpha = P_{rc}/\rho_c$) related with the relativistic limit of the Lane-Emden equation ($\alpha \to 0$ corresponds to the Newtonian limit). Note that for each polytrope model (given by $n$), $\eta$ is an increasing function that grows until the numerical evaluation stops at the surface of the compact object ($\xi_\Sigma$), where naturally coincides with the total Schwarzschild mass $M$.

\begin{figure}[ht!]
		\centering
		\includegraphics[width=0.5\textwidth]{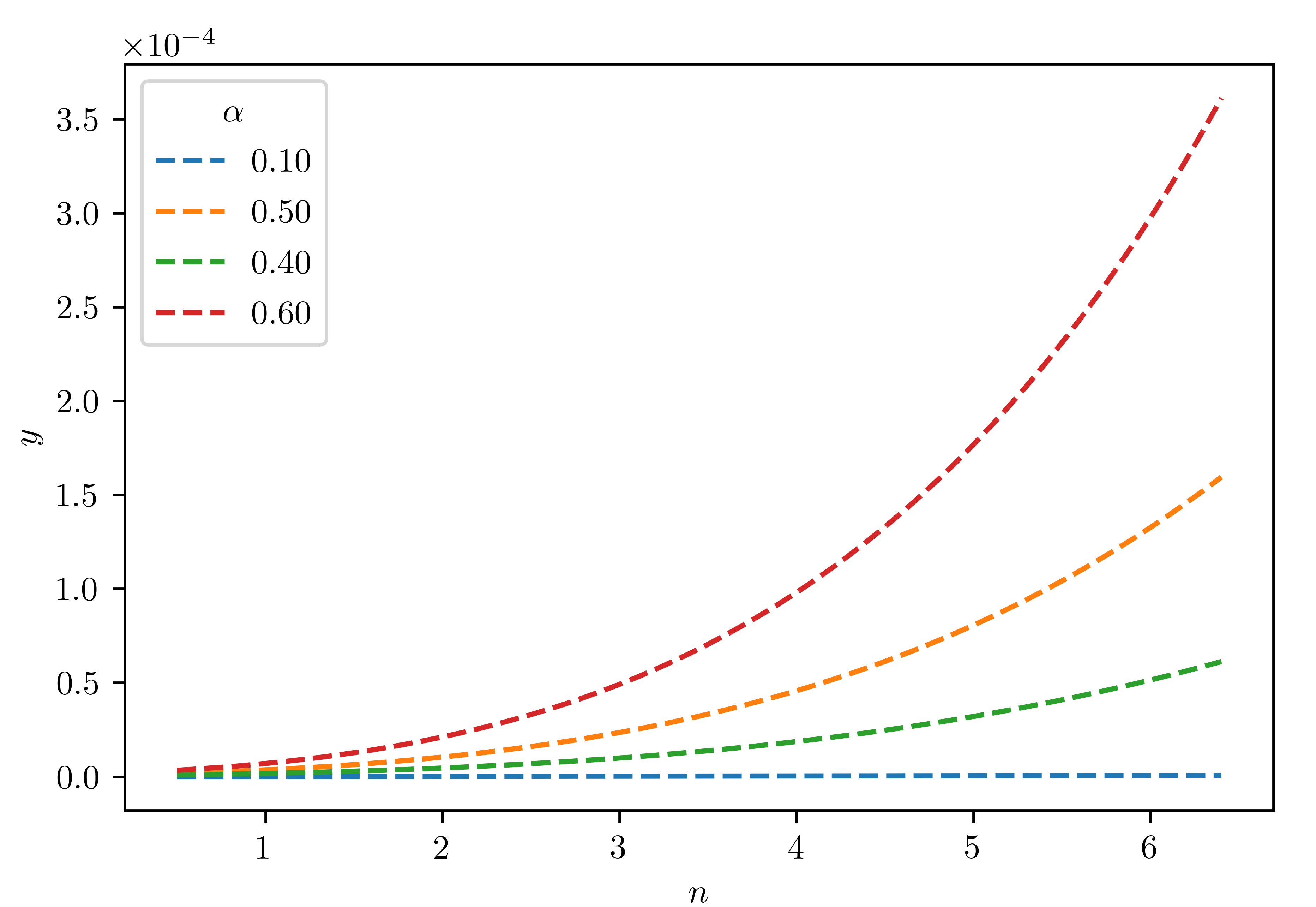}
		\caption{Surface potential $y$ as a function of $n$ for different values of $\alpha$.}
		\label{fig:tolman-y}
\end{figure}

In Fig. \ref{fig:tolman-y} its shown the ``surface potential`` $y$ which measures the degree of compactness as a function of the polytropic index $n$ for different values of $\alpha$. The parameter $y$ is a relevant variable since it measures the compactness of the configuration, it is related with the redshift  and it will also be useful in the analysis of the behavior of the active gravitational mass (Tolman-mass).

\begin{figure}[ht!]
		\centering
		\includegraphics[width=0.5\textwidth]{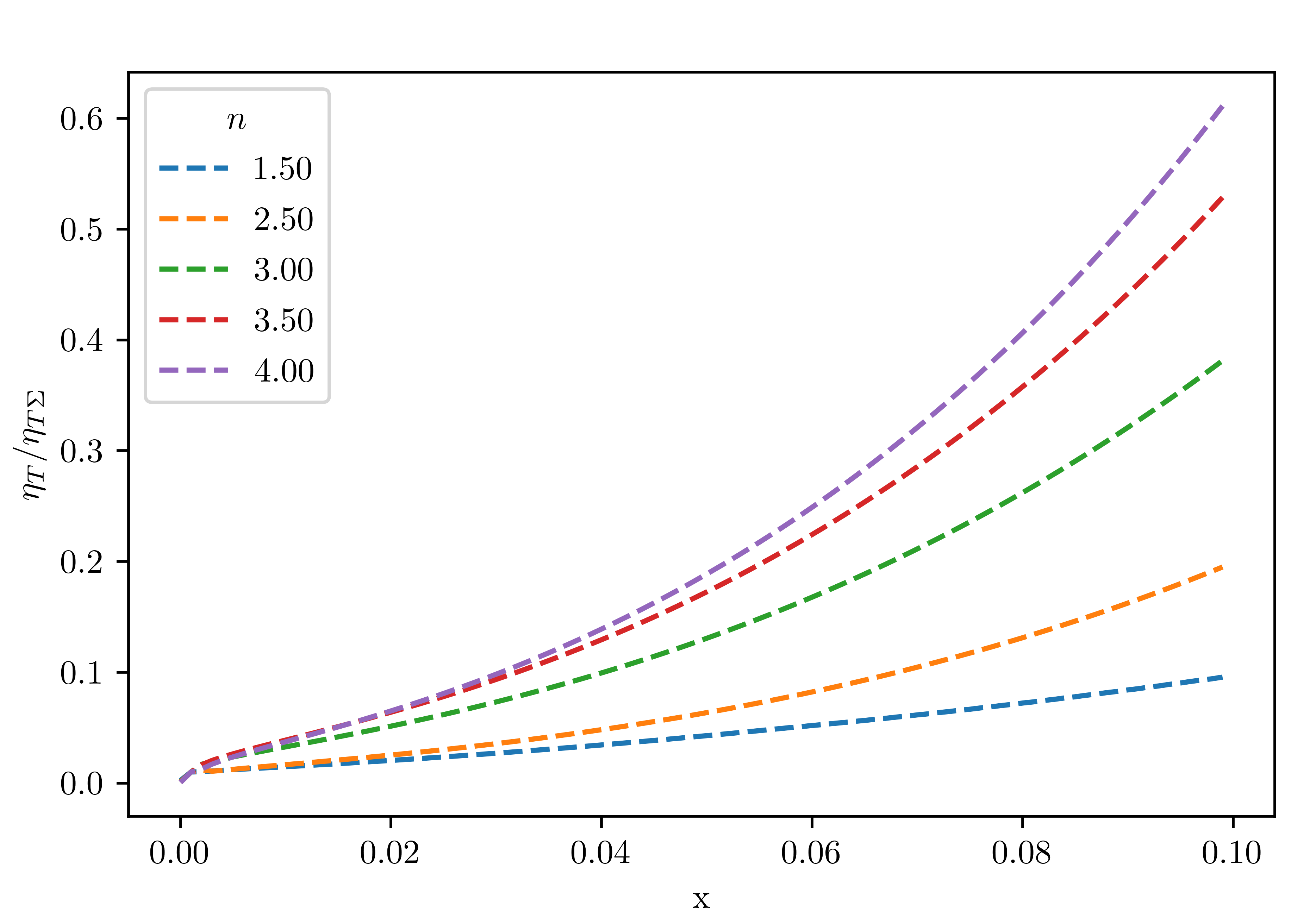}
		\caption{Normalized Tolman-Whittaker mass $\eta/\eta_{\Sigma_T}$ as a function of $x$ for $\alpha=0.2$ and different values of $n$.}
		\label{fig:tolman-masa}
\end{figure}

Fig. \ref{fig:tolman-masa} displays the Tolman mass (normalized by the total mass), for the  Tolman IV seed-solution, as function of $\xi$ for the selection of values of the parameters indicated in the legend. The behaviour of the curves is qualitatively the same for a wide range of values of the parameters. Note that to produce the numerical behaviour for $\eta/\eta_{\Sigma_T}$ (represented in Fig. \ref{fig:tolman-masa}) we must feed the data of $y$ obtained from previous Fig. \ref{fig:tolman-y}. We see that by increasing the index of the polytrope, the Tolman mass generally increases and is concentrated towards the outer surface of the object. In terms of stability, and keeping in mind the physical meaning of the Tolman mass, it may be said that more stable configurations correspond to smaller values of the Tolman mass concentrated towards the center of the sphere, since these are associated to a sharper reduction of the active gravitational mass in the inner regions, thereby providing a clear physical picture of the described scenario.

\begin{figure}[ht!]
		\centering
		\includegraphics[width=0.5\textwidth]{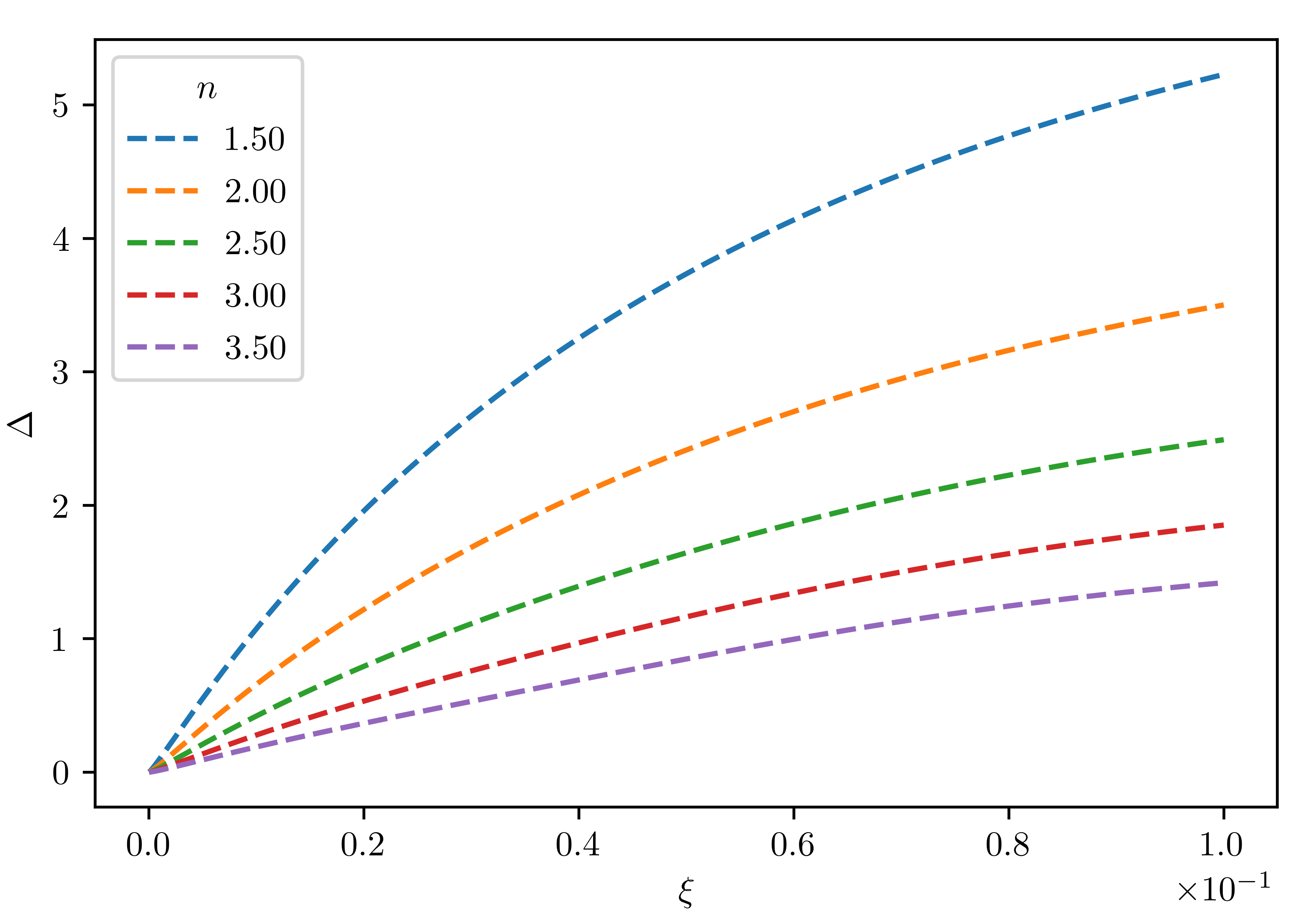}
		\caption{Local anisotropy $\Delta$ as a function of $\xi$ for $\alpha=0.5$ and different values of $n$.}
		\label{fig: tolman-ani}
\end{figure}

Fig. \ref{fig: tolman-ani} shows the dependence of the local pressure anisotropy with the dimensionless radial variable $\xi$ and the polytrope index. Note that $\Delta$ starts at zero, where $P_r = P_{\perp}$, is a monotonically increasing function as expected, and the radial pressure $P_{r}$ vanishes at the surface as required by the continuity of the second fundamental form.It is observed that the anisotropy function  decreases with the increase of the polytrope index $n$, and this fact can be related to the Tolman mass distribution, existing a relationship between the stability of compact objects and their local pressure anisotropy, an issue that has been widely reported \cite{herrera2013,6p}


\subsection{Durgapal IV}

Now, we apply the procedure described in section 4 to the Durgapal IV \cite{durgapal} interior solution as a seed in the GD framework. The starting point is
\begin{eqnarray}
e^\nu &=& a(cr^2 +1)^4\\
e^{-\mu}&=& \frac{7-10cr^2 -c^2r^4}{7(cr^2 +1)^2} + \frac{bcr^2}{(cr^2 +1)^2 (1+5cr^2)^{2/5}},\nonumber\\
\end{eqnarray}
where $a$, $b$ (with units of the inverse of a length squared) and $c$ (dimensionless) are constants. In this case, the fixed constant values used in the Durgapal IV seed are $a=0.6254$, $b=2.1479$. and $c=1.372281$.

\begin{figure}[ht!]
		\centering
		\includegraphics[width=0.5\textwidth]{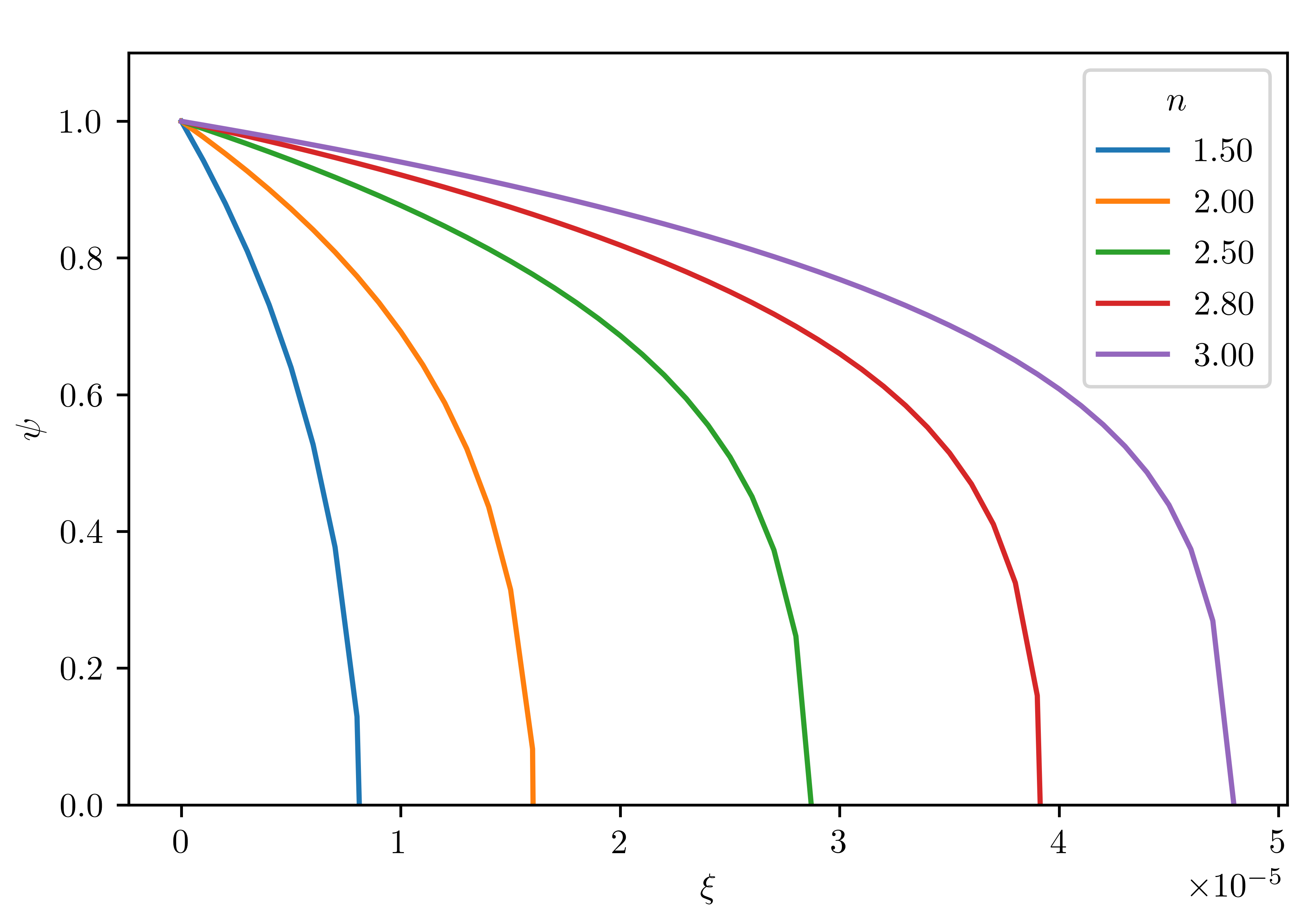}
		\caption{$\psi$ as a function of $\xi$ for $\alpha=0.5$ and different values of $n$.}
		\label{fig:durgapal-psi}
\end{figure}

\begin{figure}[ht!]
		\centering
		\includegraphics[width=0.5\textwidth]{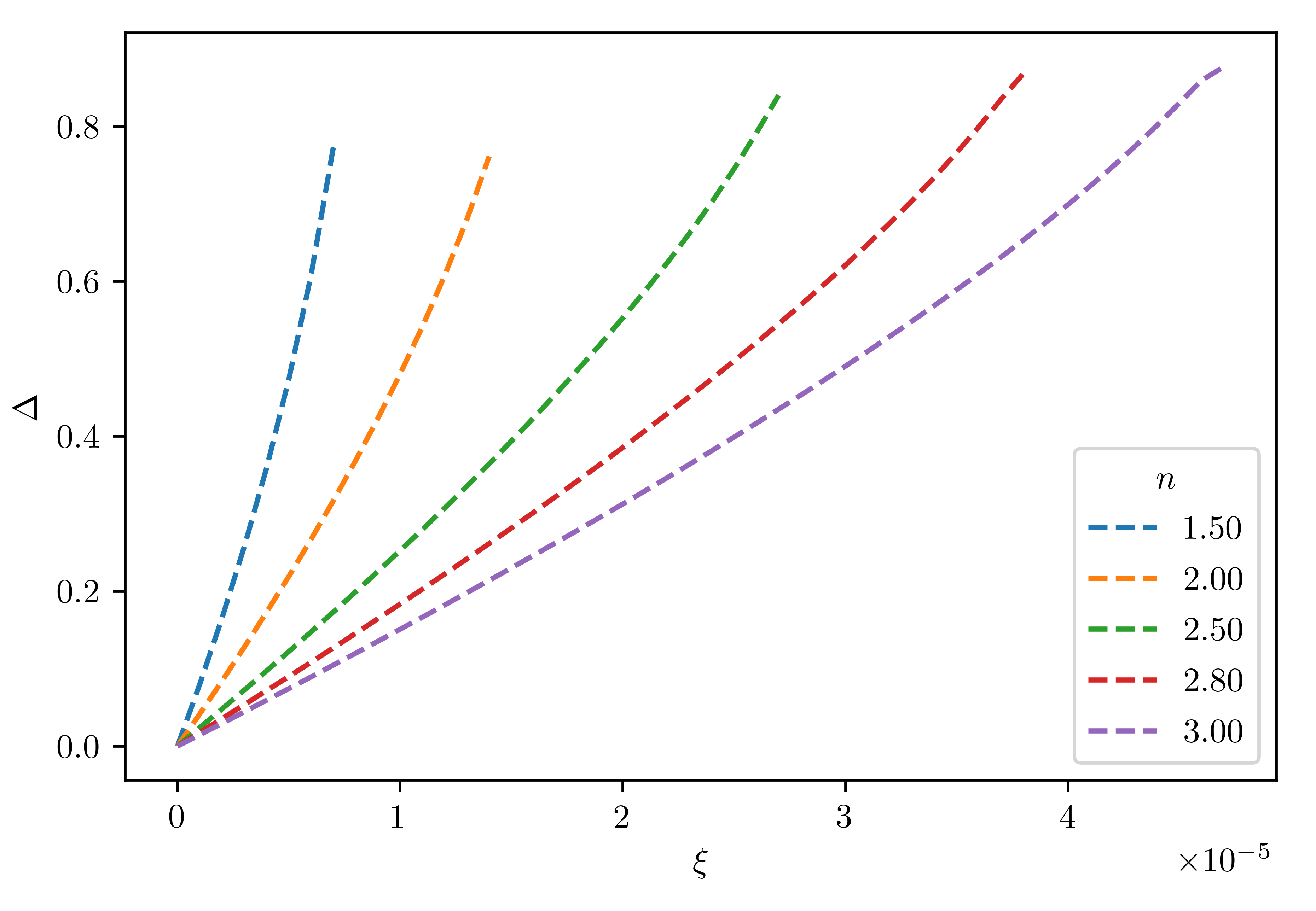}
		\caption{Local anisotropy $\Delta$ as a function of $\xi$ for $\alpha=0.5$ and different values of $n$.}
		\label{fig:durgapal-ani}
\end{figure}

In figure \ref{fig:durgapal-psi} it is shown the integration of Eqs. (\ref{lemd}) and (\ref{masseta}), with previous use of the anisotropy function (\ref{anisotropia final}), for the Durgapal IV seed-solution, plotted for the values of the parameters indicated in the figure. Note that $\psi$ is monotonously decreasing as expected. Again, the numerical analysis depends on, both, the polytropic index $n$ and the ``rigidity'' parameter $\alpha$ (related to the ratio of pressure and density at the center of the compact object). Each $n$ defines a specific stellar object that, for this model (Durgapal IV seed-solution) presents different boundaries, $\xi_{\Sigma}$, consistent with the fact that the radial pressure $P_{r}$ vanishes at the surface ($\Sigma$) as required by the continuity of the second fundamental form. However as we have mentioned, the ensuing qualitative behaviour, namely, larger values of $\psi$ (for bigger $n$) for smaller values of $\Delta$ (everywhere throughout the sphere), is maintained for a wide range of values. It is straightforward to obtain this behavior by comparing with Fig. \ref{fig:durgapal-ani}. Similar behaviors have been reported in \cite{herrera2013}.

\begin{figure}[ht!]
		\centering
		\includegraphics[width=0.5\textwidth]{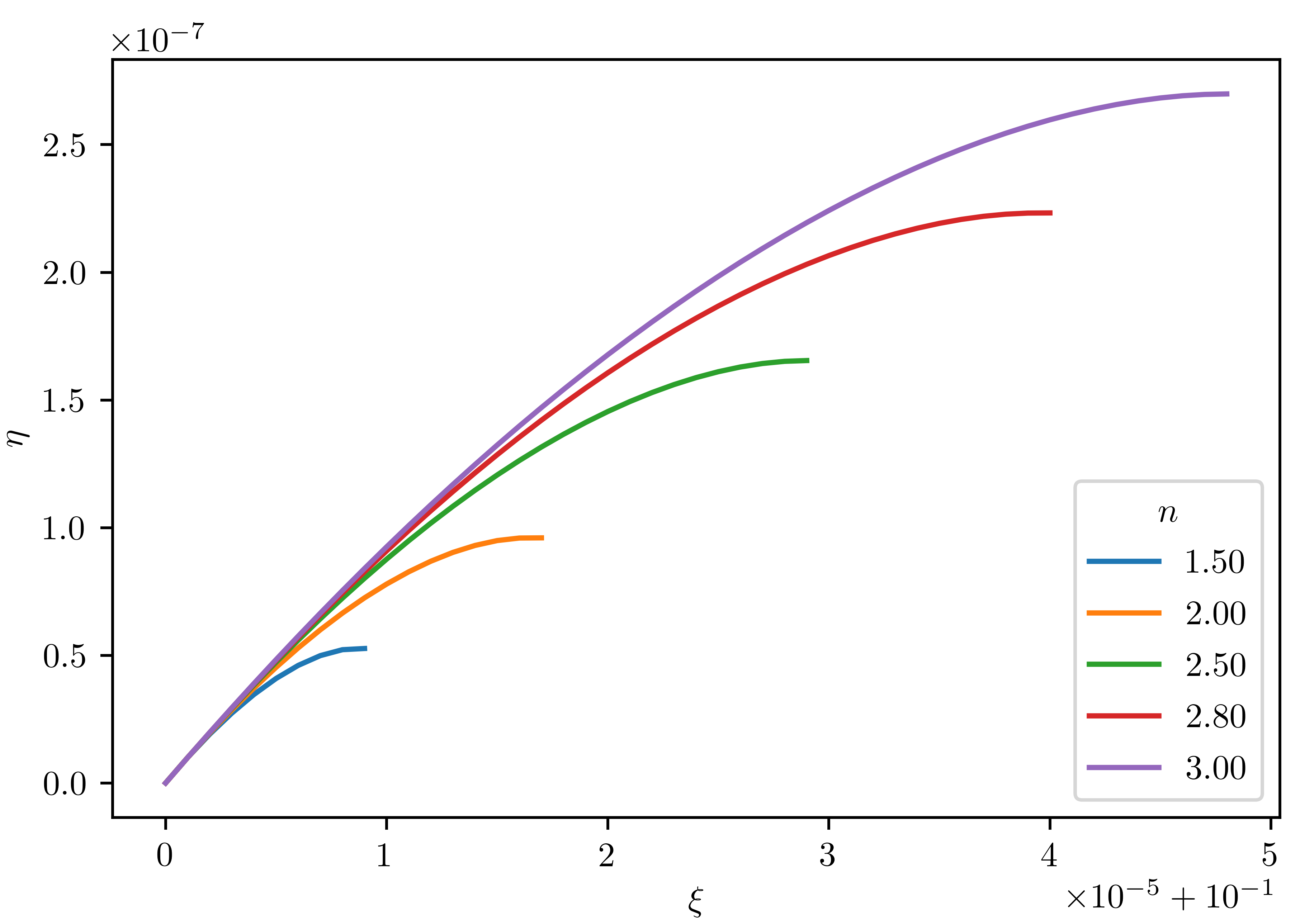}
		\caption{$\eta$ as a function of $\xi$ for $\alpha=0.5$ and different values of $n$.}
		\label{fig:durgapal-eta}
\end{figure}

The mass-function $\eta$ as function of $\xi$ for different values of $n$ is presented in Fig. \ref{fig:durgapal-eta}. The numerical evaluation has been stopped when $\eta$ reaches $\xi_\Sigma$, where the mass function, the Tolman mass, and the total mass become equal. Of course, each polytrope model, with different boundary (fact observed in Fig. \ref{fig:durgapal-psi}) in turn also present a different total contained energy. The mass function $\eta$, related with the radial metric function $\lambda$, constitutes a parameter that allows us to analyze the energy contained within the sphere, although the Tolman mass has been shown to be more useful to describe the active gravitational mass \cite{herrera2013, herrera97}.

\begin{figure}[ht!]
		\centering
		\includegraphics[width=0.5\textwidth]{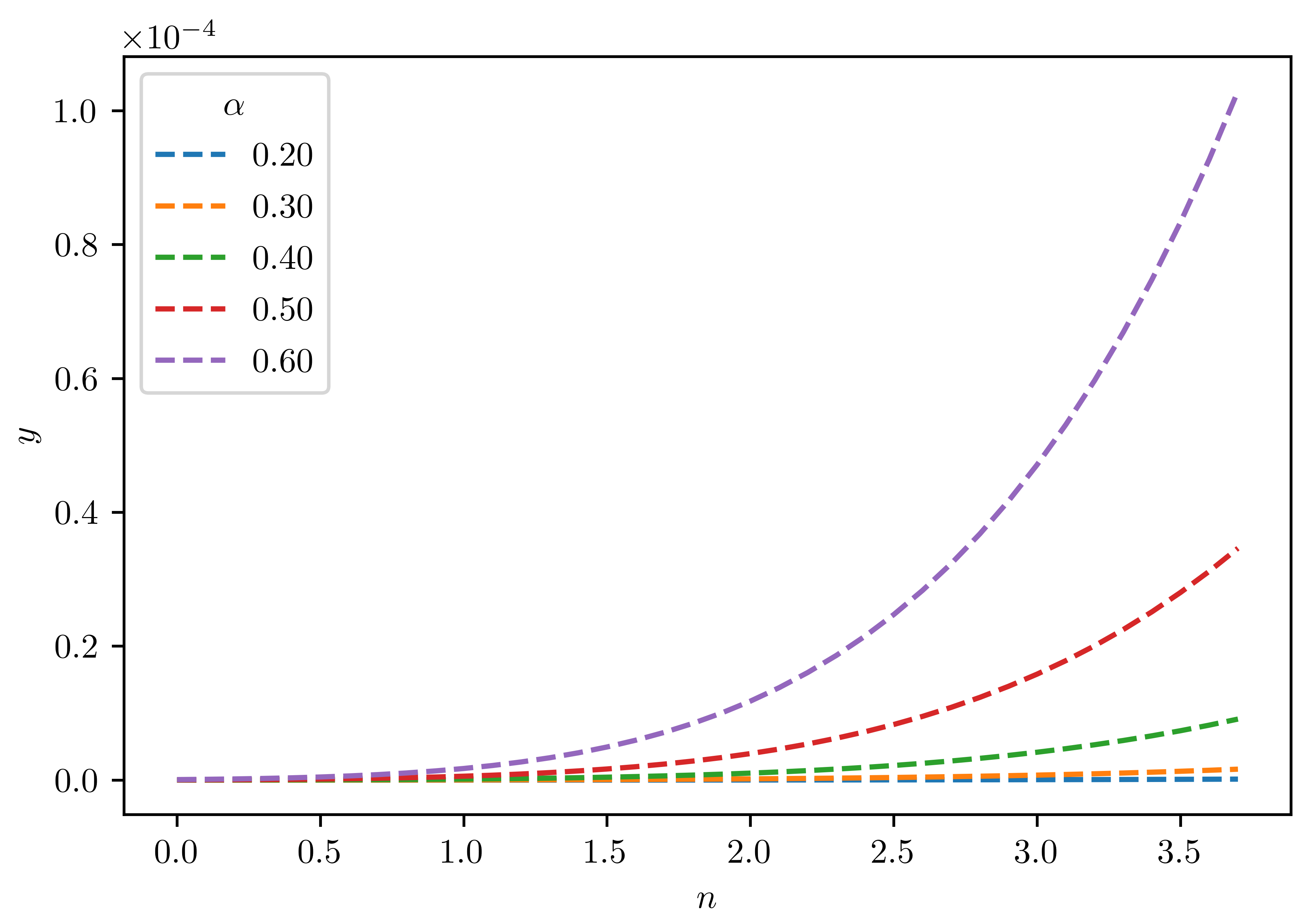}
		\caption{Surface potential $y$ as a function of $n$ for different values of $\alpha$.}
		\label{fig:durgapal-y}
\end{figure}

\begin{figure}[ht!]
		\centering
		\includegraphics[width=0.5\textwidth]{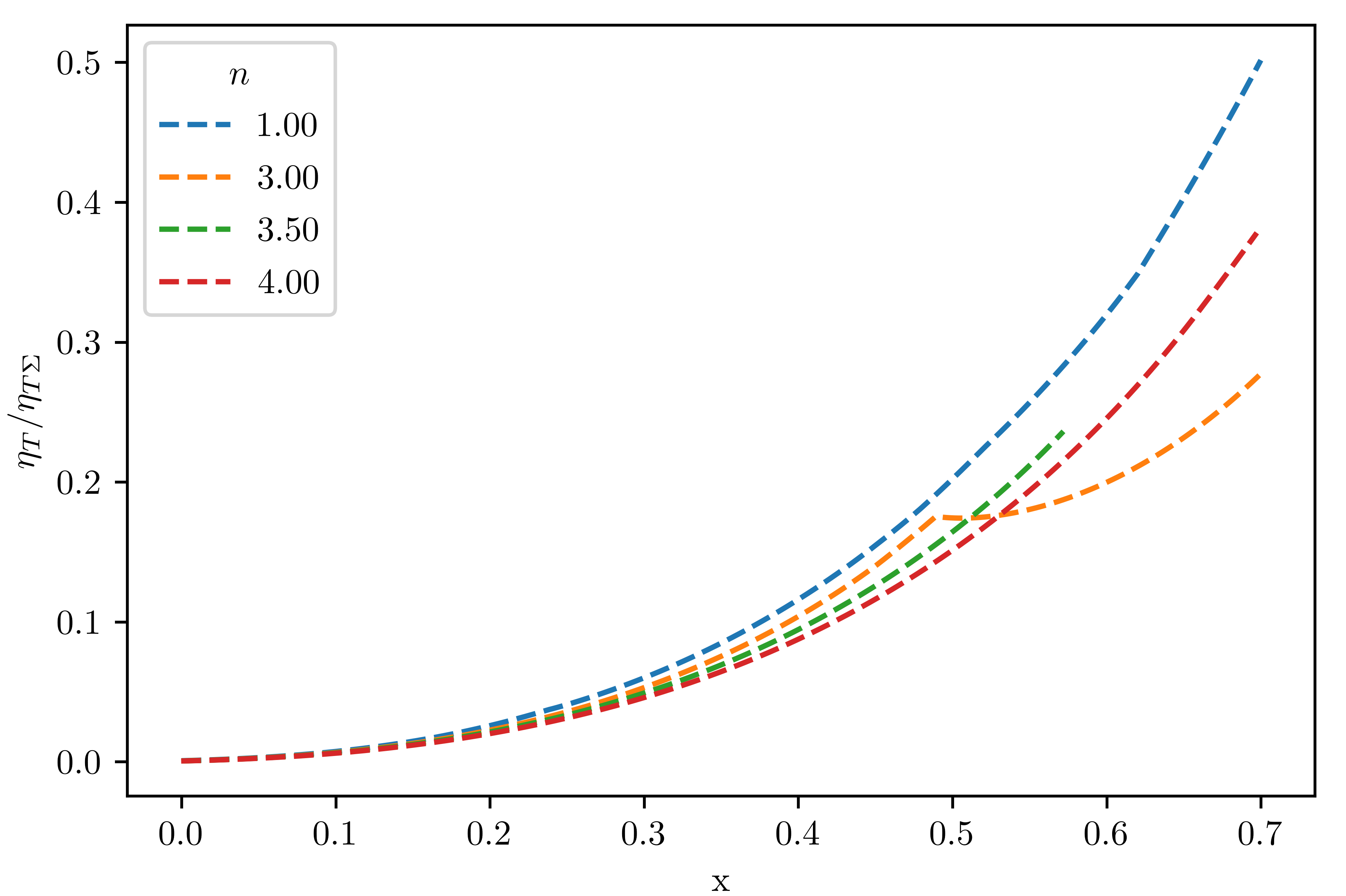}
		\caption{Normalized Tolman-Whittaker mass $\eta/\eta_{\Sigma_T}$ as a function of $x$ for $\alpha=0.5$ and different values of $n$.}
		\label{fig:durgapal-masa}
\end{figure}

The parameter $y$ is plotted in Fig. \ref{fig:durgapal-y} as function of $n$ for different $\alpha$ and the normalized Tolmann mass $\eta/\eta_{\Sigma_T}$ is shown in Fig. \ref{fig:durgapal-masa}. Again, the active gravitational mass increases from the center to the surface, showing some peculiar behaviors for some values of the polytropic index (orange line: $n=3$). This fact has been reported before (see references \cite{6p} for a more detailed discussion).

Finally, Fig. \ref{fig:durgapal-ani} is dedicated to exposing the behavior of the local anisotropy of the pressure as a function of the dimensionless parameter $\xi$. For this anisotropy function we obtain the usual behavior. We note that the anisotropy is an increasing function and when the index of the polytrope increases the anisotropy given in some layer of the stellar object decreases.


\subsection{Wyman IIa}

In this section we use the Wyman IIa solution \cite{wyman} as a seed, namely
\begin{eqnarray}
e^\nu &=& (a-br^2)^2\\
e^{-\mu}&=& 1+ cr^2(a-3br^2)^{-2/3},
\end{eqnarray}
with $a$, $b$ and $c$ are constants. For this model, the fixed constant values are $a=1.5297$, $b=1.0$. and $c=2.148$.

\begin{figure}[ht!]
		\centering
		\includegraphics[width=0.5\textwidth]{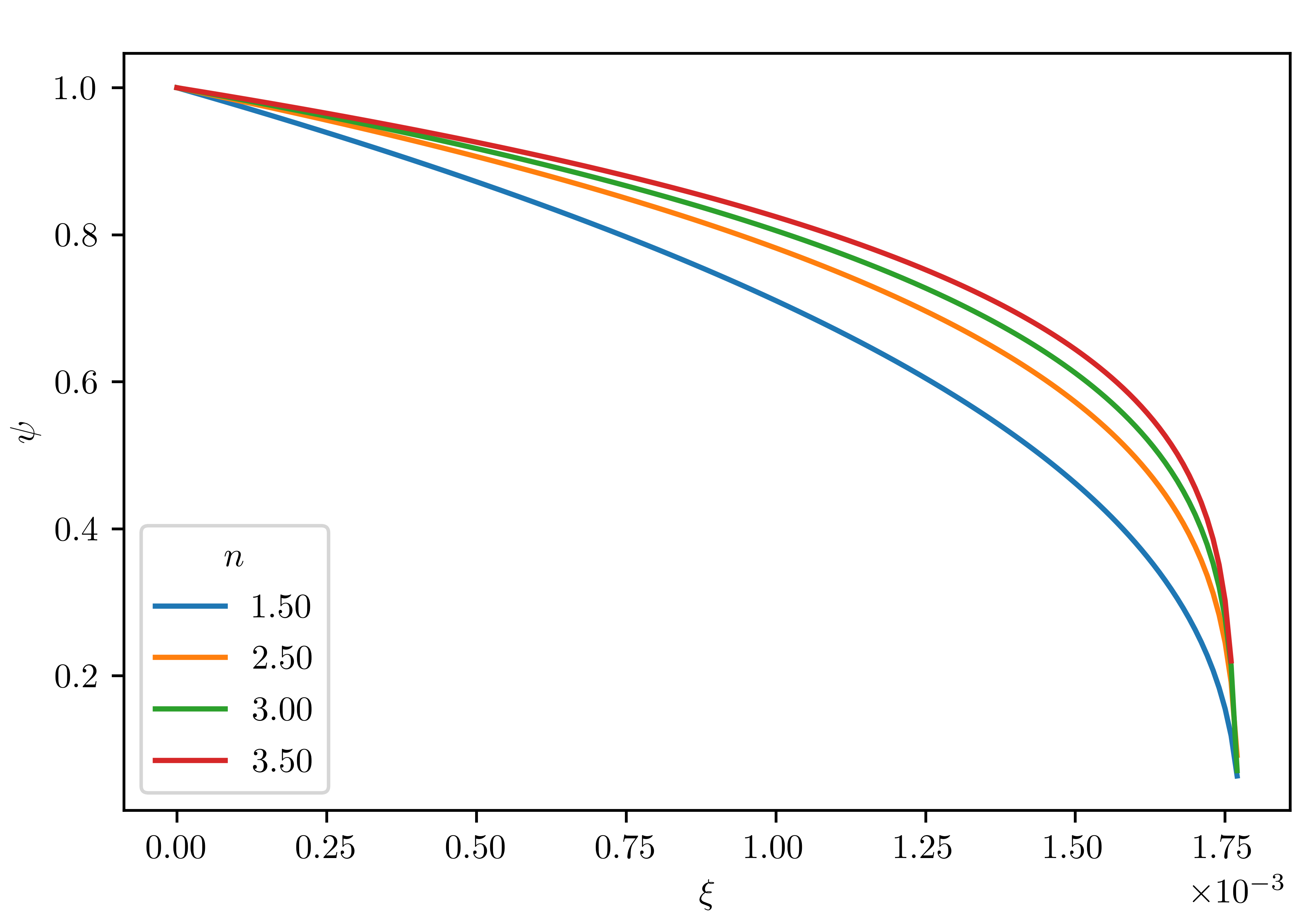}
		\caption{$\psi$ as function of $\xi$ for $\alpha=0.5$ and different values of $n$.}
			\label{fig:Wyman-psi}
\end{figure}

In Fig. \ref{fig:Wyman-psi} we show the behavior of the matter sector through integration of the Lane-Emden equation and we get $\psi$ as function of $\xi$ for the polytropic model represented by Wyman IIa seed-solution. Note that the density is positive inside the compact star, reaches its maximum at the center and decreases monotonously outwards, as expected. Unlike the two previous models, it is observed that all the configurations obtained for different values of the polytropic index, have the same boundary surface where $P_{r\Sigma}=0$.

\begin{figure}[ht!]
		\centering
		\includegraphics[width=0.5\textwidth]{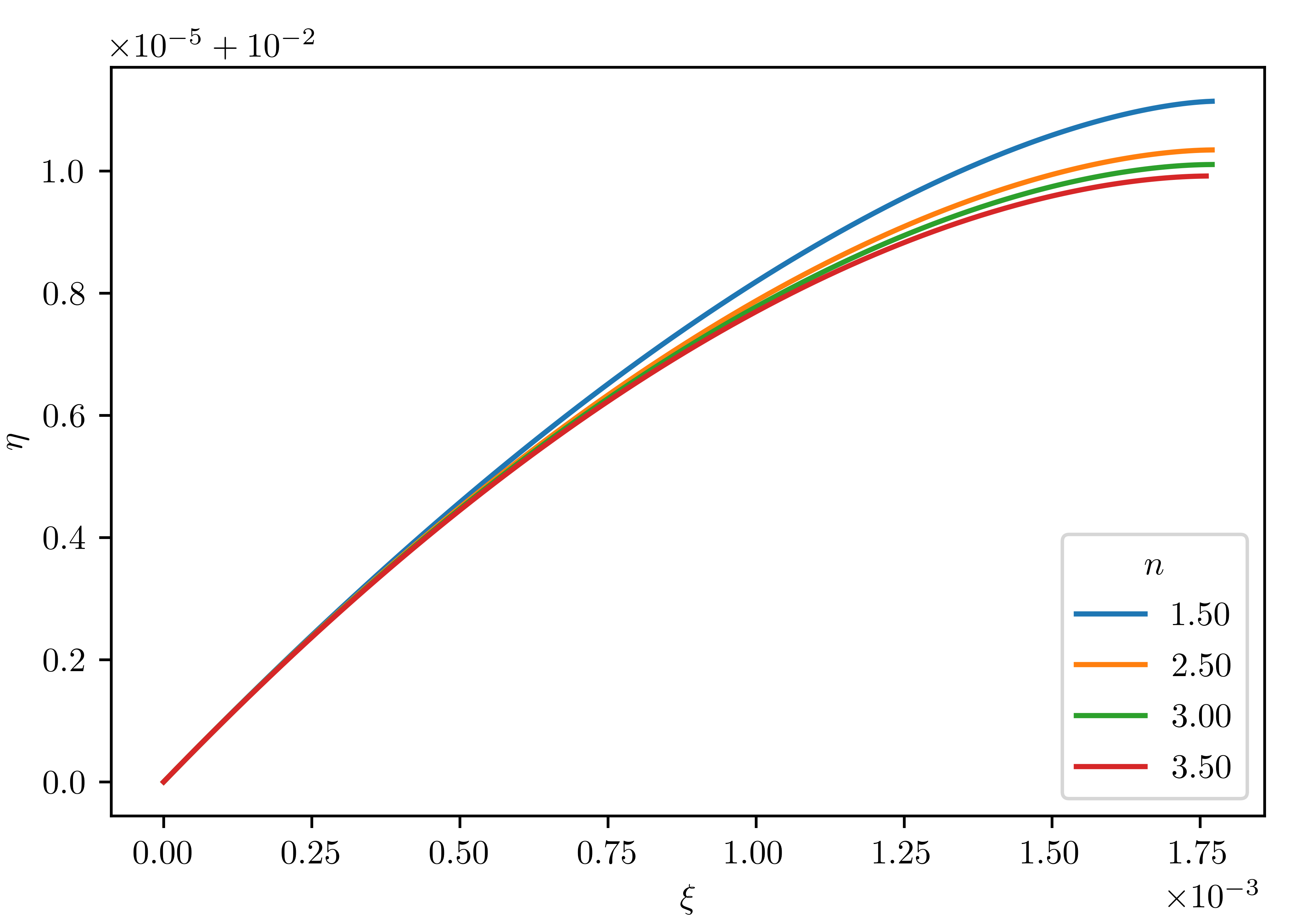}
		\caption{$\eta$ as a function of $\xi$ for $\alpha=0.5$ and different values of $n$.}
			\label{fig:Wyman-eta}
\end{figure}

In Fig. \ref{fig:Wyman-eta} we plot the $\eta$-mass as function of $\xi$ for different values of $n$ which grows appropriately from zero to its total limit value and larger values of the mass-function are obtained for smaller values of $n$. 

\begin{figure}[ht!]
		\centering
		\includegraphics[width=0.5\textwidth]{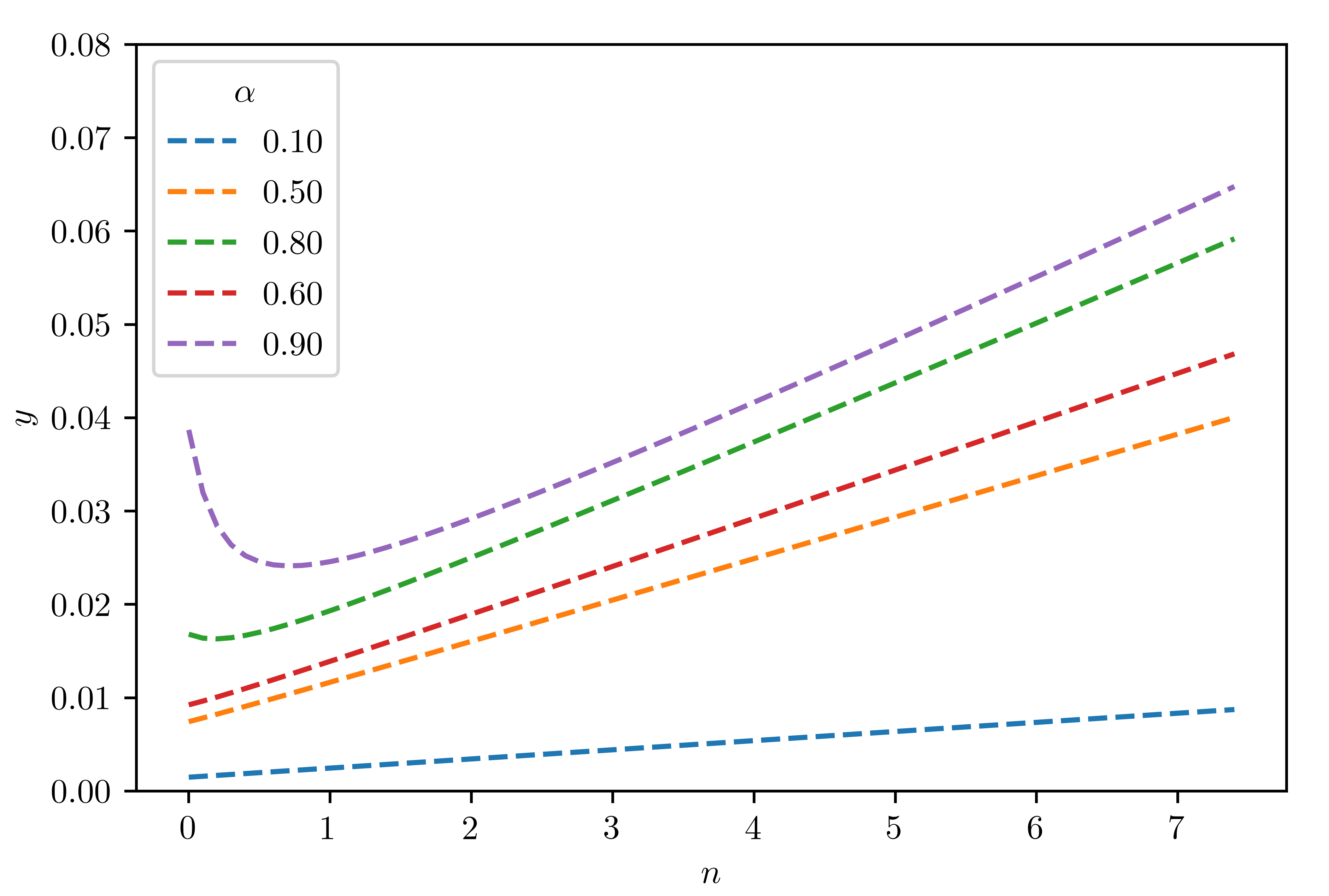}
		\caption{Surface potential $y$ as a function of $n$ for different values of $\alpha$.}
			\label{fig:Wyman-y}
\end{figure}

\begin{figure}[ht!]
		\centering
		\includegraphics[width=0.5\textwidth]{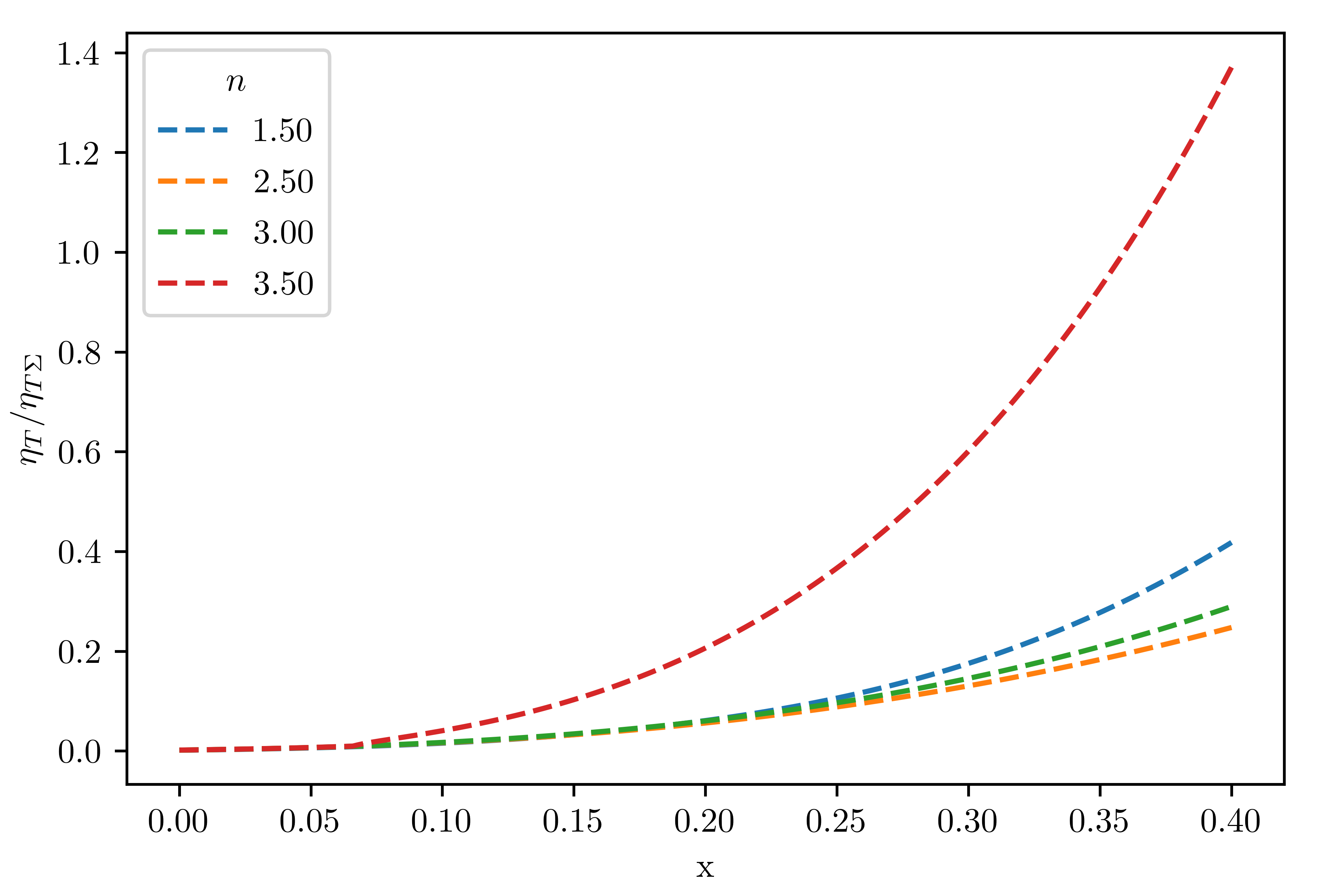}
		\caption{Normalized Tolman-Whittaker mass $\eta/\eta_{\Sigma_T}$ as a function of $x$ for $\alpha=0.7$ and different values of $n$.}
			\label{fig:Wyman-masa}
\end{figure}

The surface potential $y$ and the normalized Tolman mass, for a selection of values of the parameters are plotted in figures \ref{fig:Wyman-y} and \ref{fig:Wyman-masa} respectively. Also, the conclusions extracted from these figures are basically the same as the ones reached from the previous models. At any rate, maximal values of $y$ correspond to bigger values of $\alpha$, which represents a consistent and expected result for all models. 

\begin{figure}[ht!]
		\centering
		\includegraphics[width=0.5\textwidth]{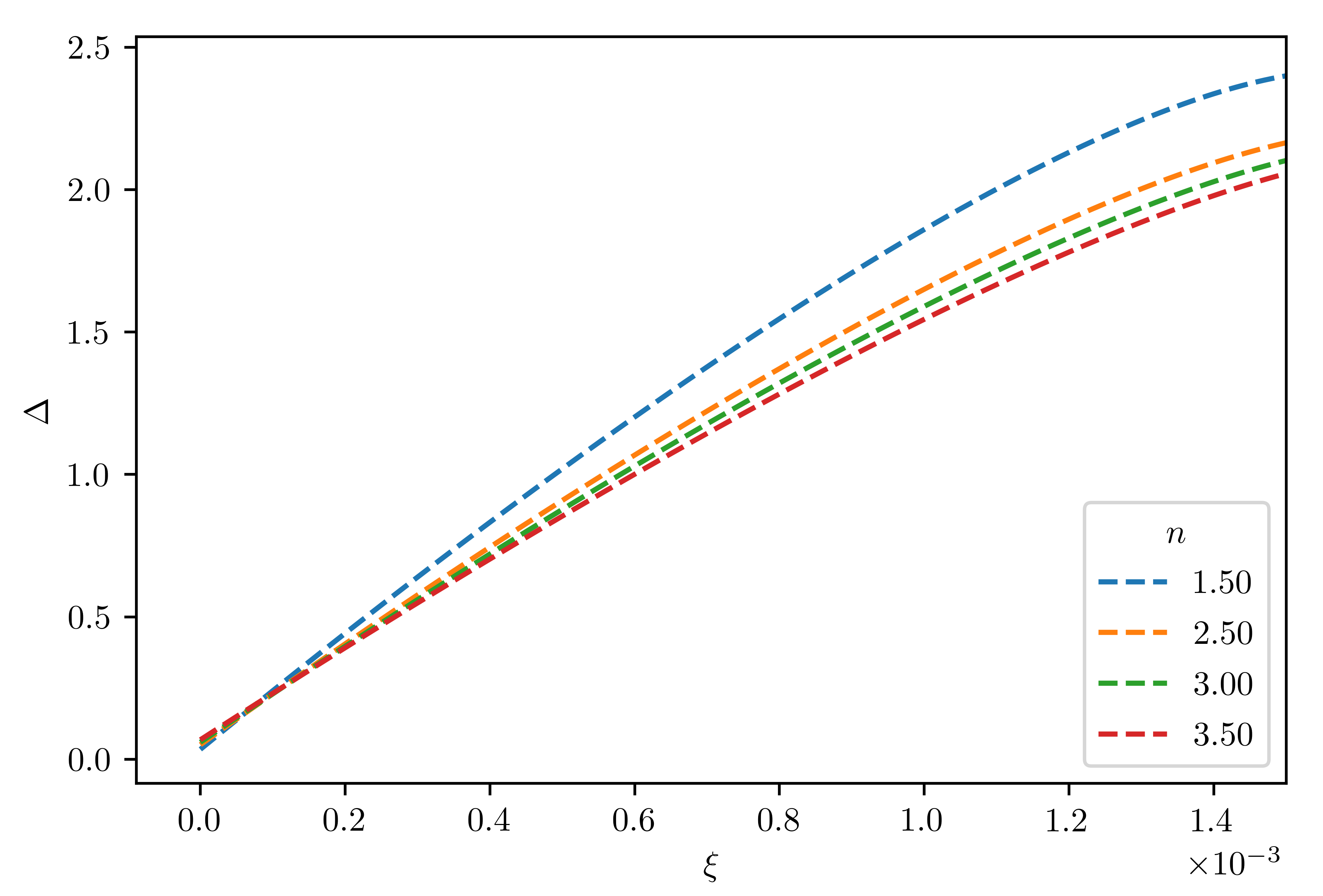}
		\caption{Local anisotropy $\Delta$ as a function of $\xi$ for $\alpha=0.5$ and different values of $n$.}
			\label{fig:Wyman-aniso}
\end{figure}

In Fig. \ref{fig:Wyman-aniso} we show the behavior of the anisotropy function $\Delta$ as a function of the redefined variable ($\xi$) for different values of the duplet of parameters ($n$, $\alpha$) indicated in the caption of the corresponding figure. An increasing behavior is manifested and the $\Delta$-function decreases with increasing $n$. In its turn, more compact configurations correspond to smaller values of the anisotropy and smaller values of the Tolman mass in the inner regions. In other words, for this case, smaller values of $n$ (corresponding to more compact configurations) reach stability by reducing the active gravitational mass in the inner regions. Therefore, it may be inferred from this figure that more stable configurations correspond to smaller values of $n$ since they are associated to a sharper reduction of the Tolman mass in the inner regions. In the same way, smaller polytrope indices imply an increase in the local pressure anisotropy, so that this function generates more stable situations for more compact spheres.



\section{Discussion}

In order to obtain a general relativistic polytrope model we must integrate the hole system of EFE. If we are assuming that the fluid pressure is anisotropic, we need to provide further information about the anisotropy inherent to the problem under consideration. The introduction of a new variable yields an additional degree of freedom and therefore Eq. (\ref{Pol}) is not enough to integrate the Lane--Emden equation. For doing that, there are different ways to deal with this problem. It can be assumed an ansatz for the anisotropy allowing a specific modeling \cite{herrera2013}. Certain physical conditions can be imposed on the metric variables such as the vanishing of the Weyl tensor (conformally flat polytrope models) \cite{prnh} and the Karmarkar Class I polytrope \cite{EF-K}. Also, the vanishing complexity polytrope model has been developed \cite{VC}.

Since polytropes represent a variety of fluid systems with a wide range of applications in astrophysics (e.g. Fermi fluids), we have described hereby a general framework for modeling general relativistic polytropes (with local pressure anisotropy), by means of the Gravitational Decoupling process, through the Minimal Geometric Deformation approach. Thus, we have built another method to obtain the generalized Lane-Emden equations, that allows us to find a specific model. Assuming the polytropic equation of state for the radial pressure allows us to obtain an expression for the total anisotropy of the system in terms of the dimensionless defined mass-function ($\eta$), given in (\ref{eq:var2}), and the metric variables of the seed sector. This represents an enormous advantage, since now we are able to obtain a great variety of general relativistic polytropes models choosing well-behaved known solutions as seeds. In this work we have studied three polytrope models considering the Tolman IV, Durgapal IV and Wyman IIa seed-solutions. For each case we have found and integrated numerically the full set of equations; the generalized Lane-Emden equations for anisotropic matter. Such extensions of the polytrope solutions to the general relativistic case are mandatory if one has to deal with ultra compact objects such as neutron stars, where general relativistic effects cannot be neglected. However it should be stressed that the above mentioned models are presented here with the sole purpose to illustrate the method. The natural way to obtain models consists in providing the specific information about the kind of anisotropy present in each problem, but having said this, it is important to emphasize that the obtained models exhibit some interesting features which deserve to be commented.

Thus, we observe in Figs. \ref{fig: tolman-psi}, \ref{fig:durgapal-psi} and \ref{fig:Wyman-psi} that bounded configurations exist for a range of values of the parameters involved and only in the Wyman IIa seed-solution model all the radii of the border coincide. However due to the existence of a larger number of parameters than in the isotropic case, the conditions for the existence of finite radius distributions are more involved than in the latter case. 

Figs. \ref{fig:tolman-y}, \ref{fig:durgapal-y} and \ref{fig:Wyman-y} show the behavior of the ``surface potential'' and for all the models considered, $y$ grows as $n$ increases and has bigger values for bigger rigidity parameter ($\alpha$). It should be emphasized that the relationship between the maximal values of $y$ (maximal surface redshift) and the local anisotropy of pressure, has been discussed in detail in the past \cite{y1, y2, y3}. The explanation for such interest is easily understood, if we recall that the surface redshift is an observable variable, which thereby might provide information about the structure of the source, that is related with the polytrope model (the index $n$) that describes the fluid and also its degree of anisotropy. This is so, since each anisotropic polytropic model is characterized by a unique $y$ as can be seen in Eq. (\ref{y}). The polytropic models favoring higher redshifts are clearly exhibited in the Fig. \ref{fig:Wyman-y}, for the Wyman IIa solution, where a peculiar behavior is observed for the highest value of $\alpha$. This  ``anomalous'' behavior could be related to the extreme (maximal) value of this parameter ($\alpha = 1$), that corresponds to the stiff equation of state $P_r=\rho$, which is believed to describe ultradense matter \cite{zh}. 

The correspondence mentioned above, between the $y$-parameter and the polytropic index $n$, suggests that a relationship can be established between compactness and the local anisotropy distribution of the fluid, which in turn is linked to interesting facts that appear in the search for the stability of the system, except for particular cases where such a correspondence is broken. 
In order to delve deeper into this feature, in Figs. \ref{fig:tolman-masa}, \ref{fig:durgapal-masa} and \ref{fig:Wyman-masa} we have investigated the behaviour of the Tolman mass (that could be related with the stability) within the sphere for each model. These facts have been extensively reported in \cite{herrera2013, prnh, 6p, herrera97, chan} where is discussed the efficiency to diminish the Tolman mass in the inner regions and to concentrate it in the outer ones depending on the anisotropic factor which brings out the role played by the anisotropy in the stability of the fluid configuration.\\ 

In principle, a somewhat speculative argument, states that it could be possible to investigate the characteristics of the fluid of the star by determining $n$ by means of the surface potential. This would be valid for such configurations where the general relativistic effects as well as the inclusion of pressure anisotropy, are unavoidable, and are modeled resorting to a polytropic equation of state. So, a potential application of the approach presented here could apply to the study of super-Chandrasekhar white dwarfs that have masses of the order of $2.8M_\odot$. Nevertheless, care must be exercised with the fact that some of the physical phenomena present in such configurations (e.g. intense magnetic fields) could break the spherical symmetry.

\end{document}